\documentstyle[openbib,12pt,fleqn]{article}
\renewcommand{\baselinestretch}{1.2}
\renewcommand{\arraystretch}{0.9}

\newcommand{\tr}{{\rm tr}\, }

\newcommand{\Lop}{{\cal L}}
\newcommand{\sym}{{s}}
\newcommand{\nsym}{{n_s}}
\newcommand{\asym}{{a}}

\newcommand{\symf}{{\tilde s}}
\newcommand{\nsymf}{n_{\tilde s}}
 
\newcommand{\fg}[1]{\item {\label{#1}}}

\newcommand{\rf}[1]{\cite{#1}}

\newcommand{\barr}{\begin{array}}
\newcommand{\bea}{\begin{eqnarray}}
\newcommand{\beq}{\begin{equation}}
\newcommand{\ear}{\end{array}}
\newcommand{\eea}{\end{eqnarray}}
\newcommand{\ceq}{\nonumber \\ & & }
\newcommand{\eeq}{\end{equation}}

\newcommand{\continue}{\nonumber \\ }

\newcommand{\spao}[1]{\mbox{\hspace{#1}}}
\newcommand{\spav}[1]{\parbox{1mm}{\vspace*{#1}}}

\newcommand{\ssty}{\scriptstyle}
\newcommand{\sssty}{\scriptscriptstyle}

\newsavebox{\ipiu}
\newsavebox{\imen}

\sbox{\ipiu}{$\ssty i \sssty +1$}
\sbox{\imen}{$\ssty i \sssty -1$}

\newcommand{\MatrixII}[4]{
   \left[
   \begin{array}{cc}
      {#1}  &  {#2}  \\ [1ex]
      {#3}  &  {#4}
   \end{array}
   \right] }

\begin{document}

\begin{titlepage}
\spav{2cm}\\
\centering \spav{1cm}\\
{\LARGE\bf 
Symmetry Decomposition of Chaotic Dynamics
\\}
\spav{2cm}\\
{\large 
Predrag Cvitanovi\'c      
}\\
{\normalsize\em Dept. of Physics, New York University,
New York, NY 10003
         \\ }
{\normalsize\em and \\}
{\normalsize\em Niels Bohr Institute$\footnotemark
                                     \footnotetext{ permanent address }$,
Blegdamsvej 17, DK-2100 Copenhagen \O
         \\ }
{\normalsize and \\}
{\large 
Bruno Eckhardt 
}\\
{\normalsize\em Fachbereich Physik der Philipps-Universit\"at,
\\Renthof 6, D-3550 Marburg   \\ }
\spav{2mm}
\vfill
{\small\bf Abstract\\}
\spav{2mm}\\
{\small\parbox{13cm}{\spao{4mm}
Discrete symmetries of dynamical flows give rise to relations between periodic
orbits, reduce the dynamics to a fundamental domain, and lead to factorizations
of zeta functions. These factorizations in turn reduce the labor and improve
the convergence of cycle expansions for classical and quantum spectra
associated with the flow. In this paper the general formalism is developed,
with the $N$-disk pinball model used as a concrete example and
a series of physically interesting cases worked out in detail.
}\\}
\spav{4mm}\\
\end{titlepage}
\setcounter{footnote}{0}

Nonlinearity 1993, to appear
\section{Introduction}
\label{Intr}
submitted to Nonlinearity Feb 26, 92;
revised Oct 19, 92;
file printed \today

The periodic orbit theory of classical chaotic dynamical systems has a long
and distinguished history; initiated by Poincar\'e\rf{poincare}, and
developed as a mathematical theory of hyperbolic dynamical systems by
Smale, Sinai, Bowen, Ruelle and others\rf{smale,sinai,bowen,ruelle},
it has in recent years been applied to many systems of physical
interest\rf{GOYcycles,CGP,GJP,eck}. The periodic orbit theory of quantum
mechanical systems largely parallels this development; originating in the
work of Hadamard\rf{Had} and Selberg\rf{selberg}, it has been developed
as a quantum mechanical theory by Gutzwiller, Balian and Bloch, Berry and
others\rf{gutz,gutz_hyp,gutbook,bb,berry78}, and has been in the focus of much
recent research. In both the classical and the quantum contexts, one is
interested in computing spectra of certain evolution operators; this can be
done by determining zeros of
Fredholm determinants or associated zeta
functions\rf{ruelle,selberg,voros}.
The periodic orbits emerge in this context essentially through the
identity $log$~$det$~=~$tr$~$log$ which relates the spectrum to the
traces of the evolution operators, {\em i.e.}, the periodic orbits or cycles.

As demonstrated in a series of papers \cite{eck},\cite{AACI}-\cite{russ},
cycle expansions of zeta functions can be profitably used for the
calculation of such spectra in chaotic dynamical systems.
These systems often come equipped with discrete symmetries, such as the
reflection and the rotation symmetries of various potentials. We shall show
here that such symmetries simplify and improve the cycle expansions in a
rather beautiful way; they can be exploited to relate classes of periodic
orbits and factorize zeta functions, not only in quantum mechanics (where the
utility of discrete symmetry factorizations is well known\rf{gasp}),
but also in classical mechanics.
The group-theoretic factorizations of zeta functions that we develop here
were first introduced and applied in ref.~\cite{eck}.
They are closely related to the symmetrizations introduced by
Gutzwiller\rf{gutz_sym} in the context of the semiclassical periodic orbit
trace formulas, recently put into  more general group-theoretic context by
Robbins\rf{robb}, whose exposition, together with Lauritzen's\rf{laur} treatment
of the boundary orbits, has influenced the presentation given here.
A related group-theoretic decomposition in context of hyperbolic
billiards was utilized in ref.~\cite{BV}.

Invariance of a system under symmetries means that the symmetry image of a
cycle is again a cycle, with the same weight. The new orbit may be
topologically distinct (in which case it contributes to the multiplicity of
the cycle) or it may be the same cycle, shifted in time. In the latter case,
the cycle can be subdivided into segments, each of which is a symmetry image
of an irreducible segment.
The period or the action of the full orbit is given by the sum along the
segments, whereas the stability is given by the product
of the stability matrices of the individual segments. The phase space can
be completely tiled by a fundamental domain and its symmetry images
(``phase space" in this paper stands for the coordinates of any $d$-dimensional
dynamical system whose evolution is described by a set of first order
differential equations or iterative mappings, not necessarily Hamiltonian).
The irreducible segments of cycles in the full space, folded back into the
fundamental domain, are closed orbits in the reduced space.

The main point of this paper is that if the dynamics possesses a discrete
symmetry, the contribution of a cycle $p$ of multiplicity $m_p$ to a
dynamical zeta function
{\em factorizes} into a product over the $d_\alpha$-dimensional
irreducible representations $D_\alpha$ of the symmetry group,
\beq
(1-t_p)^{m_p} =
        \prod_\alpha \det
        \left( 1 -  D_\alpha(h)_{\tilde{p}} t_{\tilde{p}}
        \right)^{d_\alpha} , \quad
 t_{{p}} =  t_{\tilde{p}}^{ g/m_{{p}} }
\,\, ,
\eeq
where $t_{\tilde{p}}$ is the cycle weight evaluated on the fundamental domain,
$g$ is the dimension of the group, $h_{\tilde{p}}$ is the group element
relating the fundamental domain cycle $\tilde{p}$ to a segment of the full
space cycle $p$, and $m_p$ is the multiplicity of the $p$ cycle.  Emergence
of symmetrized subspaces, a common phenomenon in quantum mechanics,
is perhaps surprising in a classical dynamics context. The basic idea is simple:
in classical dynamics, just as in quantum mechanics, the symmetrized subspaces
can be probed by linear operators (observables) of different symmetries. If a
linear operator commutes with the
symmetry, it can be block-diagonalized and the associated determinants will
therefore factorize.

This paper is meant to serve as a detailed guide to computation of zeta
functions for systems with discrete symmetries. We develop here the cycle
expansions needed for evaluation of factorized zeta functions,
and examplify them by working out a series of cases of physical interest:
$ C_2, C_{2v}, C_{3v}$ and $C_{4v}$ symmetries.
For instance, one has
a $C_2$ symmetry in the Lorenz system\rf{lor,GO}, the Ising model,
and in the 3-dimensional anisotropic Kepler
potential\rf{gutz_sym,TW,CC92},
a $C_{3v}$ symmetry in H\'enon-Heiles type potentials\rf{HH,JS,rich,laur},
a $C_{4v}$ symmetry in quartic oscillators\rf{EHP,MWR},
in the pure $x^2 y^2$ potential\rf{Mat,CP} and
in hydrogen in a magnetic field\rf{EW1},
and a $C_{2v} =C_2\times C_2$ symmetry in the stadium billiard\rf{robb}.

We will illustrate our results using the pinball scattering by three and four
disks\rf{tdisk} as an example. Besides their intrinsic interest as examples of
classical and quantum mechanical chaotic dynamics\rf{gasp,eck,steiner,CES},
they are also relevant to smooth potentials. The pinball model may be thought
of as the infinite potential wall limit of a smooth potential, and, much as the
1-d tent map captures the topology of a general unimodal map, the $N$-disk
symbol dynamics can serve as a $covering$ symbolic dynamics in smooth
potentials. One may either define potential wall collisions in
phase space\rf{EW1,EW2} or one may start with the
infinite wall limit and continuously relax an unstable cycle onto the
corresponding one for the potential under investigation. If things go well,
the cycle will remain unstable and isolated, no new orbits (unaccounted for by
the $N$-disk symbolic dynamics) will be born, and the lost orbits will be
accounted for by a set of pruning rules.  For example, this adiabatic approach
has been profitably used in ref.~\cite{DR_prl} in disproving the conjecture
that the $x^2 y^2$ potential is ergodic.  Its validity has to be checked
carefully in each application, as things can easily go wrong; for example,
near bifurcations the same naive symbol string assignments can refer to a whole
island of  distinct periodic orbits.

In addition to the symmetries exploited here, time reversal symmetry and a
variety of other non-trivial discrete symmetries can induce further relations
among orbits; we shall point out several
of examples of cycle degeneracies under time reversal.
We do not know whether such symmetries can be exploited
for further improvements of cycle expansions.

The paper is organized as follows. In the next section, we recall some basic
facts of the zeta-function formalism and cycle expansions, and describe
the symbolic dynamics of the $N$-disk model.
In section~\ref{Counti} we illustrate the utility of zeta function formalism by
using it to count cycles.
In section~\ref{degene} we introduce discrete symmetries and apply them to
identify degenerate classes of cycles.
In section~\ref{Dynami} we describe the reduction of the $N$-disk dynamics to
the fundamental domain and the special treatment required by boundary orbits.
Finally, in section~\ref{Factor} we apply symmetries to reduce the symbolic
dynamics and to factorize zeta functions. Several examples are worked out in
detail. Cycle expansions for the (symmetry unreduced) 3- and 4-disk
dynamics are listed in the appendix.
We conclude with a summary and an outlook for further work.

\section{Preliminaries}
\label{Prelim}

Here and in sect.~\ref{Counti} we review the cycle expansion
formalism; the subject proper of the paper, group-theoretic
factorizations, commences only in sect.~\ref{degene}. The reader
might profitably start with sect.~\ref{degene}, and refer back to
the preliminaries of sects.~\ref{Prelim} and~\ref{Counti} as
need arises.

\subsection{Zeta functions}

Transfer operators and the associated zeta
functions are treated extensively in the literature\rf{ruelle}.
Here we merely state the results needed for the purposes of
this paper, in the notation of ref.~\cite{AACI}.

The general setting is as follows: given a dynamical system,
presented either as a $d$-dimensional map $f(x)$, or as a $d$+$1$ dimensional
flow (in the latter case, the flow can be reduced to
a $d$-dimensional mapping by means of appropriate Poincar{\'e} sections),
one is interested in time evolution of certain distributions,
such as classical probability distributions and quantum mechanical
wave functions. The effect of the dynamics on such distributions is
given by linear evolution operators,
such as the integral kernel used in evaluation of the escape rate from a
repeller\rf{KT} described by a map $x_{n+1} = f(x_n)$,
\beq
{\cal L}(y,x)
= \delta (y - f(x))
\,\, .
\label{L_xy}
\eeq
The time dependence of distributions in question is determined by
the eigenspectrum and eigenfunctions of evolution operators, and the problem
that concerns us here is the problem of effective evaluation  of such
eigenspectra for a given dynamical system.

The eigenvalues $\lambda_k$ of ${\cal L}$ are the
inverses of the zeros of
\beq
Z(z) = \mbox{\rm Det}(1-z{\cal L}) = \prod_{k=0}^\infty (1-z \lambda_k)
\,\, .
\label{hadamard}
\eeq
The periodic orbit approach to the determination of zeros of $Z(z)$
is based on the observation that the determinant of an
operator is  related to its traces by
\beq
\det(1-z{\cal L}) =
  \exp\left( {-\sum_{n=1}^{\infty} \frac{z^n}{n}\,\tr{\cal L}^n }\right)
\label{det_tr}
\eeq
Traces of an operator like  (\ref{L_xy}) receive contributions from
all prime cycles $p$ of period $n_p$
and their multiple traversals, weighted by the cycle Jacobians
\beq
Z(z) =
{\rm exp}  \left( - {
         \sum_{p} \sum_{r=1}^\infty {1 \over r}
 {  z^{ n_p r }
 \over  { | \det \left( {\bf 1}-{\bf J}_p^{r} \right) | } }
         } \right)
\,\,  .
\label{det(1-L)}
\eeq
This expression  
for $\mbox{\rm Det}(1-z{\cal L}) $ is specific to the mapping (\ref{L_xy}).
In more general settings\rf{ruelle}, $z^{ n_p}$ is replaced by a weight
whose precise form depends on the particular average being computed;
for example, in the corresponding determinant for classical
smooth flows, $z^{ n_p}$ is replaced by $e^{s T_p}$, where
${T_p}$ is the $p$ cycle period\rf{CE}.
The quantum mechanical kernel corresponding to
(\ref{L_xy}) is smeared out by a path integral, but
in the semiclassical approximation\rf{gutbook} the zeta
function\rf{voros} has essentially the same form as (\ref{det(1-L)}):
\[
Z(E) =
{\rm exp}  \left( - {
         \sum_{p} \sum_{r=1}^\infty {1 \over r}
 {  e^{\frac{i}{\hbar} S_p(E) r - i  \pi \mu_p r/2}
 \over  { | \det \left( {\bf 1}-{\bf J}_p^{r} \right) |^{1/2} } }
         } \right)
\,\,  .
\]\noindent
$S_p$ denotes the classical action and $\mu_p$ is the Maslov index.
As the group-theoretic factorizations that we shall develop here rely only on
the linearity of evolution operators, they will apply to both the classical
and the quantum cases.

For evaluation of spectra, the expansion (\ref{det(1-L)}) can be used
pretty much as it stands\rf{CPR}, it can be expanded as a multinomial in cycle
weights\rf{losal}, or the leading eigenvalues can be extracted from the
associated dynamical zeta function\rf{ruelle}
\beq
1 / \zeta =
\prod_p  ( 1 -   t_p) , \quad t_p = z^{n_p}/\Lambda_p
\quad \mbox{or} \quad
t_p = e^{\frac{i}{\hbar} S_p(E) - i  \pi \mu_p /2}/\Lambda_p^{1/2}
\,\, ,
\label{zeta__}
\eeq
obtained by approximating $\det \left( {\bf 1}-{\bf J}_p \right) =
\prod_{a=1}^d \left( 1-\Lambda_{p,a} \right)$
by the product of expanding eigenvalues $ \Lambda_p=|\prod_a^{e}\Lambda_{p,a}|$
in (\ref{det(1-L)}).
For example, for a 1-dimensional expanding map
each periodic point contributes with
the weight $1/|1-\Lambda_p^r|$:
\[
     \sum_{n=1}^\infty { z^n \over n} tr{\cal L}^n
           = \sum_p \sum_{r=1}^\infty {1 \over r}
       {z^{n_p r} \over |1-\Lambda_p^r|}
\, .
\]\noindent
Substituting
$
1/|1-\Lambda| = |\Lambda|^{-1} (1-\Lambda^{-1})^{-1}
= |\Lambda|^{-1} \sum \Lambda^{-k}
$
we obtain
\bea
\det(1-z{\cal L}) &=&
                     \exp\left( -
     \sum_p \sum_{k=0}^\infty \sum_{r=1}^\infty
      {1 \over r}
       \left({ z^{n_p} \over {|\Lambda_p| \Lambda_p^{k} } }\right)^{r}
                        \right)
                                                \nonumber\\
                  &=&
                  \prod_p \prod_{k=0}^\infty
          \left(1- {t_p \over \Lambda_p^k} \right)
\, , \quad \quad \quad
          t_p = { z^{n_p} \over {|\Lambda_p|} } \, .
\label{fredh}
\eea
The dynamical zeta function (\ref{zeta__}) is the $k=0$ part of
the above Selberg-type product.
The full zeta functions (\ref{det(1-L)})
are infinite products over dynamical zeta functions
which depend on both the expanding and
the contracting eigenvalues. Most of the developments
below are independent of the precise form of the zeta functions.

As the dynamical zeta functions have particularly simple cycle expansions,
a simple geometrical shadowing interpretation of their convergence,
and as they suffice for determination of leading eigenvalues, we
shall concentrate in this paper on their factorizations; the
full $Z(z)$ determinants can be factorized by the same techniques.
To emphasize the group theoretic structure of zeta functions, we shall
combine all the non-group-theory dependence of a
$p$-cycle into a cycle weight $t_p$.
We shall also often absorb $z$ into the transfer operator:
$z{\cal L} \rightarrow {\cal L} $, $z^{n_p} t_p \rightarrow t_p$.

The first prerequisite for converting expressions like (\ref{zeta__})
into cycle expansions is efficient enumeration of periodic orbits - the problem
to which we turn next.

\subsection{Symbolic dynamics}

The key to a theory of a chaotic dynamical system is its qualitative,
topological description\rf{smale}, or, as it is usually called, its
{\em symbolic dynamics}. The strategy is to partition phase space into
topologically distinct regions, associate with each region a symbol from an
{\em alphabet}, and use those symbols to label every possible trajectory.
{\em Covering} symbolic dynamics assigns a distinct label to each
distinct trajectory, though there might be symbol sequences which are not
realized as trajectories. If all possible symbol sequences can be realized as
physical trajectories, the symbolic dynamics is called {\em complete};
if some sequences are not allowed, the symbolic dynamics is {\em pruned}
(the word is suggested by ``pruning'' of branches corresponding to forbidden
sequences for symbolic dynamics organized
hierarchically into a tree structure). In that case
the alphabet must be supplemented by a {\em grammar},
a set of pruning rules.
A periodic symbol string corresponds to a periodic orbit or a {\em cycle}.
Periodic orbits will here be distinguished by a bar over the primitive symbol
block, but we often omit the bar if it is clear from the context that we are
dealing with a periodic orbit.

These concepts are easily illustrated by the pinball models that we shall study
here. Consider the motion of a point particle in a plane with $N$ elastically
reflecting convex disks. Any trajectory can be labelled by the
sequence of disk bounces, and a symbolic dynamics is given by the alphabet of
$N$ symbols $\{1,2,3,\cdots,N\}$. As the bodies are convex, there can be no two
consecutive reflections off the same disk, hence the first rule of the
grammar of allowed sequences is
that symbol repetitions $\_ 11 \_ $, $\_ 22\_$, $\cdots $, $\_ NN \_$
are forbidden.  More generally, we shall refer to a symbolic dynamics
as an ``$N$-disk" symbolic dynamics if the phase space can be
partitioned in $N$ distinct regions such that an orbit starting in a
partition can in one step reach all other partitions except itself.

A finite length scattering trajectory is not uniquely specified by its
(finite) symbol sequence, but an unstable cycle (consisting of infinitely many
repetitions of a prime building block) is. We shall show in
sect.~\ref{Counti} that the prime cycles for
such simple grammars are easily enumerated; for example, table 1 contains a
list of all prime 3-disk cycles up to length 9, and table 2 contains a list of
prime 4-disk cycles.

An important effect of a discrete symmetry is that
it tesselates the phase space into copies
of a fundamental domain, and thus induces a natural partition of phase
space. The group elements $g = \{a,b,\cdots,d\}$ which
map the fundamental domain $\tilde{M}$ into its copies $g \tilde{M}$, can double
in function as letters of a symbolic dynamics alphabet.
If the dynamics is symmetric under interchanges of disks,
the absolute disk labels $\epsilon_i=1,2,\cdots,N$ can
be replaced by the symmetry-invariant relative disk$\rightarrow$disk increments
$g_i$,
where $g_i$ is the discrete group element that maps disk $i-1$ into disk $i$.
Experience shows that more often
than not specifics of the model at hand dictate the choice of symmetry
reduced symbolic dynamics, so rather than attempting to
develop a general procedure here, we shall demonstrate the reduction for a
series of specific examples in sect.~\ref{Factor}.
An immediate gain arising from symmetry invariant relabelling
is that $N$-disk symbolic dynamics becomes ($N-1$)-nary,
with no restrictions on the allowed sequences\rf{tdisk,grass}.
However,
the main gain is in the close connection between the symbol string symmetries
and the phase space symmetries which will aid us in the zeta function
factorizations. Once the connection between the full space and the reduced
space is established, working in the fundamental domain ({\em ie}.,
with irreducible segments) is so much simpler that we
never use the full space orbits in actual computations.

Whether this symbolic dynamics is complete (as is the case for sufficiently
separated disks), pruned (for example, for touching disks), or only a
first coarse graining of the topology (as, for example, in systems with
islands of stability) depends on the details of the dynamics and requires
further case-by-case investigation. In the $N$-disk model outlined above we
have tacitly assumed that the disks are sufficiently separated so that all
possible symbol sequences (excluding symbol repeats) are realized as physical
trajectories. If the disks shadow each other, further infinite families of
sequences are pruned\rf{bunim,PS,hansen}.
Determining the pruning rules is in general a
highly non-trivial undertaking, carried out so far for only a few dynamical
systems\rf{CGP,AGIP}. In this paper we assume that the disks are sufficiently
separated so that there is no additional pruning beyond self bounces. This
assumption does not affect our main result, the symmetry induced factorizations
-- it only affects the cycle counting of sect.~\ref{Counti}.

\subsection{Cycle expansions}

In the simplest of the cases that we shall discuss here (an
example is the fundamental
domain symmetric subspace of a 3-disk repeller, discussed in
sect.~\ref{C3v_fact}) the system is described by a complete binary symbolic
dynamics.
The Euler product (\ref{zeta__}) is given by\rf{AACI}
\bea
1/\zeta & = & (1-zt_{0})(1-zt_{1})(1-z^2 t_{01})(1-z^3 t_{001})(1-z^3 t_{011})
\ceq
            (1-z^4t_{0001})
            (1-z^4t_{0011})(1-z^4t_{0111})(1-z^5t_{00001})
            (1-z^5t_{00011}) \ceq
            (1-z^5t_{00101})(1-z^5t_{00111})(1-z^5t_{01011})(1-z^5t_{01111})
\dots
\label{zetabin}
\eea

The cycle expansion is obtained by multiplying out the Euler product and
grouping together the terms of the same total string length (same power of z):
\bea
1/\zeta &= &  1 - zt_0 - zt_1
 - z^2\lbrack (t_{01} - t_1 t_0) \rbrack    \ceq
 - z^3\lbrack (t_{001}- t_{01} t_0)
 - (t_{011}- t_{01} t_1) \rbrack \ceq
 - z^4\lbrack (t_{0001}  - t_0 t_{001}) + (t_{0111} - t_{011} t_{1}) \ceq
\quad  + (t_{0011}  -  t_{001} t_{1} - t_{0} t_{011} + t_0 t_{01} t_1) \rbrack-
\dots
\label{curvbin}
\eea

The terms grouped in brackets are the {\em curvature} corrections\rf{cycprl};
the terms grouped in parenthesis are combinations of longer orbits and their
shorter ``shadowing" approximants. In the counting limit $t_p=1$, and all
such shadowing combinations vanish (as we show in more detail in
sect.~\ref{Counti}). The practical utility of cycle expansions,
in contrast to direct averages over periodic orbits such as in the
trace formula\rf{gutz}, lies precisely in this organization into nearly
cancelling combinations: cycle expansions are dominated by short cycles, with
long cycles giving exponentially decaying corrections.

Further examples of cycle expansions are given in sect.~\ref{Factor}
and in the Appendix.

\section {Counting cycles}
\label{Counti}

In this section we apply the cycle-counting methods of ref.~\cite{AACI} to
the $N$-disk problem. This section is not essential to the main,
group-theoretic thrust of this paper, but in practice cycle counting is
of some use as a check of correctness of various cycle expansions.

There are $N_n=N^n$ possible distinct strings of length $n$ composed of $N$
letters. These $N_n$ strings include all $M_d$ prime cycles whose period $d$
equals or divides $n$, and each $d-$cycle contributes its cyclic permutations,
$d$ in number:
\beq
 N_n\,=\,\sum _{d|n}\,d M_d \, .
\eeq
The number of prime cycles follows by M\"obius inversion\rf{hardy}
\beq
M_n\,=\,n^{-1}\,\sum_{d|n}\,\mu \left( {{n}\over {d}}\right)\, N_d \,.
\label{Moeb_inv}
\eeq
where $\mu (1) = 1$, $\mu (n) = 0$ if $n$ has a squared factor, and
$\mu (p_1 p_2 \dots p_k ) = (-1)^k $ if all prime factors are different.

For example, from two symbols $\{0,1\}$ one can form
$M_n = 2, 1, 2, 3, 6, 9, 18,  
\dots $ prime cycles,
$i.e.$, there are two fixed points $\overline{0}$ and
$\overline{1}$, one prime 2-cycle $\overline{01}$, two 3-cycles
$\overline{001}$ and $\overline{011}$, three 4-cycles
$\overline{0001}$, $\overline{0011}$, $\overline{0111}$
(note that {\em e.g.} $\overline{1010} = \overline{01}^2$ is not prime),
and so forth. In table 3 we list the number of prime orbits up to length 10
for 2-, 3- and 4-letter complete symbolic dynamics.

For a generic dynamical system not all $M_n$ prime periodic symbol strings
are realized as physical orbits: in such cases $M_n$ is only an upper bound
to the actual number of prime $n$-cycles. To count correctly, we need to
$prune$ the disallowed orbits, $ie$. specify the grammar of the allowed
sequences. A simple example of pruning is the exclusion of ``self-bounces''
in the $N$-disk pinball. To determine the number of periodic orbits, consider
a $[N \times N]$ transfer matrix whose elements are $T_{ij}=1$ if a transition
from disk $j$ to disk $i$ is possible, and $0$ otherwise. The number of points
that are mapped back onto themselves after $n$ iterations is given by
$N_n = {\rm tr\ } T^n$. For a complete $N$-ary dynamics all entries
would equal unity
\beq
T_c = \pmatrix{ 1 & 1 & \ldots & 1 \cr
                1 & 1 & \ldots & 1 \cr
            \vdots & \vdots & \ddots & \vdots \cr
                1 & 1 & \ldots & 1 \cr }
\,\, ,
\eeq
and ${\rm tr\ } T_c^n = N^n$. The pruning of self-bounces eliminates the
diagonal entries, $T_{N-disk} = T_c - {\bf 1}$, so the number of the
$N$-disk periodic points is
\beq
N_n = {\rm tr\ }  T_{N-disk}^n  = (N-1)^n + (-1)^n (N-1) \, .
\label{N_n}
\eeq
M\"obius inversion (\ref{Moeb_inv}) now yields
\bea
M_n^{N-disk} &=&
{1 \over n}\sum_{d|n}\,\mu \left( {{n}\over {d}}\right)\, (N-1)^d
+ {{N-1} \over n}\sum_{d|n}\,\mu \left( {{n}\over {d}}\right)\, (-1)^d
\continue
             &=& M_n^{(N-1)} \quad \mbox{\rm for} \quad n>2
\,\, .
\label{mndisk}
\eea
There are no fixed points, $M_1^{N-disk}=0$. The number of periodic points
of period 2 is
$N^2-N$, hence there are $M_2^{N-disk}=N(N-1)/2$ prime cycles
of length 2;
for lengths $n>2$, the number of prime cycles is the same
as for the complete ($N-1$)-ary dynamics.

The simplest application of the cycle expansion of dynamical zeta function
(\ref{zeta__}) is the evaluation of the {\em topological entropy}. The
topological entropy is the growth rate of the number of orbits as a function
of the length of their symbol sequences:
\beq
h
 = \lim_{n \rightarrow \infty} {{\ln N_n} / n}
\,\, .
\eeq
The topological entropy
is given by the logarithm of the largest eigenvalue of the transition matrix
$T$ such as the one given in the above example. Dynamical zeta functions enter
via the relationship (see eq.~(18) of ref.~\cite{AACI})
\beq
\det (1- z T) = \prod_p (1-t_p) \, ,
\eeq
where for the topological entropy the weight assigned to a prime cycle $p$ of
length $n_p$ is $t_p = z^{n_p}$ if the cycle is allowed, or $t_p = 0$ if it is
pruned. Expanded in powers of $z$ one finds
\beq
\zeta(z)^{-1}\,=\,
\prod_{p} \,{\left(1\,-\,
t_p \right)}\,
=\,1+ \sum_{k=1}^{\infty} z^k\,c_k \, .
\label{z_pol}
\eeq
This function is called the topological zeta function\rf{smale,art-maz};
if the grammar is finite, it reduces to the topological polynomial. The
topological entropy $h$ is given by the smallest zero $z=e^{-h}$.

For complete symbolic dynamics of $N$ symbols, the topological polynomial
is simply
\beq
\zeta (z)^{-1} = { 1-Nz } \, ,
\label{ztop}
\eeq
whence the topological entropy $h = \ln N$.

One consequence of the finitness of topological polynomials is that
the contributions to curvatures at every order are even in number, half with
positive and half with negative sign. For instance, for complete binary
labelling (\ref{curvbin}),
\bea
c_4 & = & -t_{0001}\,-\,t_{0011}\,-\,
t_{0111}\,-\,t_0 t_{01} t_1       \ceq
\,+\,t_0 t_{001}\,+\, t_0 t_{011}
\,+\,t_{001} t_1\,+\, t_{011} t_1 \, .
\label{c4}
\eea
We see that $2^3$ terms contribute to $c_4$, and exactly half of them appear
with a negative sign - hence if all binary strings are allowed, this term
vanishes in the counting expression. The number of terms can be counted
using the identity
\beq
\prod_{p}\left(1+t_p \right)
=\prod_{p}
{{1-{t_p}^2} \over { 1-t_p }} \, .
\label{ccount}
\eeq
Substituting (\ref{ztop}) we obtain
\beq
\prod_{p}\left(1+t_p \right)
={{1-Nz^2} \over { 1-Nz }}=
1+Nz+\sum_{k=2}^{\infty} z^k\left(N^k-N^{k-1} \right) \, .
\eeq
The $z^n$ coefficient in the above expansion is the number of terms
contributing to $c_n$, so we find that for complete symbolic dynamics of
$N$ symbols and $n>1$, the number of terms contributing to $c_n$ is
$(N-1) N^{n-1}$.

This technique is easily generalized to cycle expansions for the N-disk
symbol sequences. Consider for example the 3-disk pinball. The prohibition of
repeating a symbol affects counting only for the fixed points and the 2-cycles.
Everything else is the same as counting for a complete binary
dynamics (eq~(\ref{mndisk})). To
obtain the topological zeta function, just divide out the binary 1- and
2-cycles $(1-zt_0)(1-zt_1)(1-z^2t_{01})$ and multiply with the correct 3-disk
2-cycles $(1-z^2 t_{12})(1-z^2 t_{13})(1-z^2 t_{23})$:
\bea
1/\zeta_{3-disk} &=& (1-2z) {(1-z^2)^3 \over (1-z)^2 (1-z^2)}
        \continue
&=& (1-2z)(1+z)^2 = 1 - 3 z^2 - 2 z^3
\> .
\label{3disk_top}
\eea
The 4-disk pinball topological polynomial can be derived in the same way: the
pruning affects again only the fixed points and the 2-cycles
\bea
1/\zeta_{4-disk} &=& (1-3z) {(1-z^2)^6 \over (1-z)^3 (1-z^2)^3}
        \continue
&=& (1-3z)(1+z)^3 = 1 - 6 z^2 - 8 z^3 - 3 z^4
\,\, ,
\label{4disk_top}
\eea
and, more generally, for an $N$-disk pinball, the topological polynominal is
given by
\bea
1/\zeta_{N-disk} &=& \left(1-(N-1)z\right)
        {(1-z^2)^{N(N-1)/2} \over { (1-z)^{N-1} (1-z^2)^{(N-1)(N-2)/2}} }
                        \continue
                 &=& \left(1-(N-1)z\right)(1+z)^{N-1}
\,\, .
\label{Ndisk_top}
\eea
The topological polynomial has a root $z^{-1}=N-1$, as we already know it
should from (\ref{N_n}).
We shall see in sect.~\ref{Factor} that the other roots
reflect the symmetry factorizations of zeta functions.
Since the zeta functions reduce to polynomials,
we are assured that there are just a few
fundamental cycles and that all long cycles can be grouped into curvature
combinations. For example, the fundamental cycles in (\ref{3disk_top}) are the
three $2$-cycles which bounce back and forth between two disks and the two
$3$-cycles which visit every disk. It is only after these fundamental
cycles have been included that a cycle expansion is expected to start
converging smoothly, $i.e.$, only for $n$ larger than the lengths of the
fundamental cycles are the curvatures $c_n$, a measure of the deviations
between long orbits and their short cycle approximants, expected to fall off
rapidly with $n$.

Conversely, if the grammar is not finite and there is no
finite topological polynomial, there will be no ``curvature'' expansions, and
the convergence will be poor. That is the generic case, and
one strategy\rf{AACI,hansen,AGIP} for dealing with it
is to find a good sequence of approximate but finite grammars;
for each approximate grammar cycle expansions yield
exponentialy accurate eigenvalues, with succesive approximate grammars
converging toward the desired infinite grammar system.

The $N$-disk topological zeta function may also be used to count the
number of terms in the curvatures.  For example, for the 3-disk pinball we get
\beq
\prod_{p}\left(1+t_p \right)
 ={ { 1-3 z^4-2 z^6} \over {1-3 z^2-2 z^3} }
=  1+3 z^2+2 z^3 + {{z^4 (6 +12 z +2 z^2) } \over {1-3 z^2-2 z^3} }
\eeq
The coefficients are 1, 0, 3, 2, 6, 12, 20, 48, 84, 184, $\dots$. That
means that, {\em e.g.}, $c_6$ has a total of 20 terms, in agreement
with the explicit 3-disk cycle expansion (\ref{3dzeta}) of the
appendix.

The above concludes our review of cycle expansions for general dynamical
systems; now we turn to the main subject of this paper, the role of
discrete symmetries in cycle expansions.

\section{Discrete symmetries}
\label{degene}

A dynamical system is invariant under a symmetry group
$G=\{{ e},{ g}_2,\ldots,{ g}_g\}$ if the equations of motion are invariant
under all symmetries ${ g} \in G$. For a map $x_{n+1} = f(x_n)$ and
the evolution operator ${\cal L}(y,x)$ defined by (\ref{L_xy}) this means
\bea
f(x) & = & {\bf g}^{-1} f({\bf g}x)
\continue
{\cal L}(y,x) &  = &  {\cal L}({\bf g}y,{\bf g}x)
\,\, .
\label{L_inv}
\eea
Bold face letters for group elements indicate a suitable representation
on phase space.
For example, if a $2$-dimensional map has the symmetry
$x_1 \rightarrow - x_1$, $x_2 \rightarrow - x_2$, the symmetry group $G$
consists of the identity and $C$, a rotation by $\pi$ around the origin.
The map $f$ must then commute with rotations by $\pi$, $f({\bf C}x)={\bf
 C}f(x)$,
with ${\bf C}$ given by the $\lbrack 2 \times 2 \rbrack$ matrix
\beq
{\bf C} =
\MatrixII{ -1} {0}{0}{ -1} \, .
\label{C_matrix}
\eeq
$C$ satisfies $C^2=e$ and can be used to decompose the phase space
into mutually orthogonal symmetric and antisymmetric
subspaces by means of projection operators
\bea
P_{A_1} &=& \frac{1}{2}({\bf e}+{\bf C} )
\,\, , \quad \quad
P_{A_2}=\frac{1}{2}({\bf e}-{\bf C} ) \, ,
     \continue
{\cal L}_{A_1}(y,x)&=&
P_{A_1} {\cal L} (y,x)=
\frac{1}{2}\left({\cal L} (y,x) + {\cal L} (-y,x)\right) \, ,
     \continue
{\cal L}_{A_2}(y,x)&=&
P_{A_2} {\cal L} (y,x)=
\frac{1}{2}\left({\cal L} (y,x) - {\cal L} (-y,x)\right)
\,\, .
\label{C_proj}
\eea
More generally\cite{robb,laur} the projection operator onto the $\alpha$
irreducible subspace of dimension $d_\alpha$ is given by
$ P_\alpha = \left(d_\alpha/|G|\right)  \sum \chi_\alpha (h) {\bf h}^{-1}$,
where $ \chi_\alpha (h) = \tr D_{\alpha}( h)$
are the group characters, and  the transfer operator ${\cal L}$ splits into
a sum of inequivalent irreducible subspace
contributions
$ \sum_\alpha \tr {\cal L}_\alpha$,
\beq
{\cal L}_\alpha(y,x) =
  \frac{d_\alpha}{|G|} \sum_{h \in G}
  \chi_\alpha (h) {\cal L} ({\bf h}^{-1} y,x)
\,\, .
\label{L_alpha}
\eeq
The prefactor $d_\alpha$ in the above reflects the fact that a
$d_\alpha$-dimensional representation occurs $d_\alpha$ times.

\subsection{Cycle degeneracies}

If ${ g} \in G$ is a symmetry of the dynamical
problem, the weight of a cycle $p$ and the weight of its
image under a symmetry transformation ${ g}$ are equal, $t_{gp} = t_p$.
The number of degenerate cycles (topologically distinct, but mapped into
each other by symmetry transformations) depends on the cycle symmetries.
Associated with a given cycle ${p}$ is a maximal subgroup
${\cal H}_{{p}} \subseteq G$, ${\cal H}_{{p}} =
\{ e, b_2, b_3, \ldots , b_h \}$ of order $h_{p}$, whose elements leave ${p}$
invariant. The elements of the quotient space $b \in G/{\cal H}_{{p}}$
generate the degenerate cycles $b {p}$, so the multiplicity of a degenerate
cycle is $m_p=g/h_{{p}}$.

Taking into account these degeneracies, the Euler product (\ref{zeta__})
takes the form
\beq
\prod_{p} (1-t_p) =
     \prod_{\hat{p}} (1-t_{\hat{p}} )^{ m_{\hat{p}} }.
 \label{eul_sym}
\eeq
Here ${\hat{p}}$ is one of the $m_p$ degenerate cycles, picked to serve
as the label for the entire class. Our labelling convention is usually
lexical, $i.e.$, we label a cycle $p$ by the cycle point whose label has the
lowest value, and we label a class of degenerate cycles
by the one with the lowest label $\hat{p}$.
In what follows we shall drop the hat in $\hat{p}$ when it is clear
from the context that we are dealing with symmetry distinct classes
of cycles.

\subsection{Example: $C_{3v}$ invariance}

An illustration of the above is afforded by $C_{3v}$, the group of symmetries
of a pinball with three equal size, equally spaced disks,
fig.~\ref{fgC3v}.
The group consists\rf{hamer} of the identity element $e$, three reflections
across axes $\{ \sigma_{12}, \sigma_{23}, \sigma_{13} \}$, and two rotations by
$2\pi/3$ and $4\pi/3$ denoted $\{ C_3, C_3^2\}$, so its dimension is $g=6$.
On the disk labels $\{1,2,3\}$ these symmetries act as permutations which map
cycles into cycles. For example, the flip across the symmetry axis going
through disk $1$ interchanges the symbols $2$ and $3$; it maps
the cycle $\overline{12123}$ into $\overline{13132}$,
fig.~\ref{fg3_exmp}a.

The subgroups of $C_{3v}$ are $C_v$, consisting of the identity and any one
of the reflections, of dimension $h=2$, and $C_3 = \{ e, C_3, C_3^2 \}$,
of dimension $h=3$, so possible cycle multiplicities are $g/h = 2$, $3$ or $6$.

The $C_3$ subgroup invariance is exemplified by the cycles $\overline{123}$
and $\overline{132}$ which are invariant under rotations by $2\pi/3$ and
$4\pi/3$, but are mapped into each other by any reflection,
fig.~\ref{fg3_exmp}b;
${\cal H}_{{p}} = \{e,C_3,C_3^2\}$, and the degeneracy is $g/h_{c_3}=2$.

The $C_v$ type of a subgroup is exemplified by the invariances of
$\hat{p}=1213$. This cycle is invariant under reflection
$ \sigma_{23} \{\overline{1213}\} = \overline{1312} = \overline{1213} $,
so the invariant subgroup is ${\cal H}_{\hat{p}}= \{ e, \sigma_{23} \}$.
Its order is $h_{C_v}=2$, so the degeneracy is $ m_{\hat{p}}= g/h_{C_v}=3$;
the cycles in this class, $\overline{1213}$, $\overline{1232}$ and
$\overline{1323}$, are related by $2\pi/3$ rotations,
fig.~\ref{fg3_exmp}c.

A cycle of no symmetry, such as $\overline{12123}$, has ${\cal H}_{{p}}=\{e\}$
and contributes in all six terms (the remaining cycles in the class are
$\overline{12132}$, $\overline{12313}$,
$\overline{12323}$, $\overline{13132}$ and $\overline{13232}$),
fig.~\ref{fg3_exmp}a.

Besides the above discrete symmetries, for Hamiltonian systems cycles may be
related by time reversal symmetry. An example\rf{freddy} are the cycles
$\overline{121212313}$ and $\overline{121212323}=\overline{313212121}$
which are related by no space symmetry (fig.~\ref{fg3_exmp}d).

The Euler product (\ref{zeta__}) for the $C_{3v}$ symmetric 3-disk problem is
given in the Appendix, eq.~(\ref{3dzeta}).

\section{Dynamics in the fundamental domain}
\label{Dynami}

So far we have used the discrete symmetry to effect a reduction in the number
of independent cycles in cycle expansions. The next step achieves much more:
the symmetries can be used to restrict all computations to a
{\em fundamental domain}. We show here that to each global cycle $p$
corresponds a fundamental domain cycle $\tilde p$. Conversely, each
fundamental domain cycle $\tilde p$ traces out a segment of the global cycle
$p$, with the end point of the cycle $\tilde p$ mapped into the irreducible
segment of $p$ with the group element $h_{\tilde p}$.

If the dynamics is invariant under a discrete symmetry, the phase space
$M$ can be completely tiled by the fundamental domain
$\tilde{M}$ and its images $a \tilde{M}$, $b \tilde{M}$, $\dots$ under the
action of the symmetry group $G=\{e,a,b,\ldots\}$,
\[
M = \sum_{a\in G} M_a = \sum_{a \in G} a \tilde{M} \, .
\]\noindent
In the above example (\ref{C_matrix}) with symmetry group
$G=\{{ e},{ C}\}$, the phase space $M=\{x_1$-$x_2$~plane$\}$
can be tiled by a fundamental domain $\tilde{M}=\{$half-plane~$x_1\geq 0\}$, and
${ \bf C}\tilde{M}=\{$half-plane~$x_1\leq 0\}$, its image under rotation by
 $\pi$.

If the space $M$ is decomposed into $g$ tiles, a function $\phi(x)$ over
$M$ splits into a $g$-dimensional vector $\phi_a(x)$ defined by
$\phi_a(x)=\phi(x)$ if $x \in M_a$, $\phi_a(x)=0$ otherwise.
Let ${h}={ab}^{-1}$ be the symmetry operation that
maps the endpoint domain $M_b$ into the starting point domain
$M_a$, and let $D(h)_{ba}$, the left regular representation, be the
$\lbrack g \times g \rbrack$ matrix whose
$b,a$-th entry equals unity if $a=hb$ and
zero otherwise; $D(h)_{ba} = \delta_{hb,a}$.
Since the symmetries act on phase space as well, the operation ${h}$ enters in
two guises: as a $\lbrack g \times g \rbrack$ matrix $D(h)$ which simply
permutes the domain labels, and as a $\lbrack d \times d \rbrack$
matrix representation ${\bf h}$ of a discrete symmetry operation on
the $d$ phase-space coordinates. For instance, in the above example
(\ref{C_matrix}) $h\in C_2$ and $D(h)$ can be
either the identity or the interchange of the two domain labels,
\beq
D(e)=\MatrixII{ 1} {0}{0}{ 1}\,\, ,
\quad
D(C)=\MatrixII{0} {1}{1}{0}
\,\, .
\label{C_mat_perm}
\eeq
Note that $D(h)$ is a permutation matrix, mapping a tile $M_a$ into
a different tile $M_{ha} \neq M_a$ if $h \neq e$. Consequently only
$D(e)$ has diagonal elements, and $\tr D(h) = g \delta_{h,e}$. However,
the phase-space transformation ${\bf h} \neq {\bf e}$ leaves invariant
sets of {\em boundary} points; for example, under reflection 
${\bf \sigma}$ across a symmetry axis, the axis itself remains invariant.
The boundary periodic orbits that belong to such point-wise invariant
sets will require special care in $\tr \Lop$ evaluations.

One can associate to the
evolution operator (\ref{L_xy}) a $\lbrack g \times g \rbrack$ matrix
evolution operator defined by
\[
{\cal L}_{ba} (y,x) = D(h)_{ba} {\cal L} (y,x)
\,\, ,
\]\noindent
if $x \in M_a$ and $y \in M_b$, and zero otherwise.
Now we can use the invariance condition (\ref{L_inv}) to move the starting
point $x$ into the fundamental domain $x={\bf a}\tilde{x}$,
${\cal L} (y,x) = {\cal L} ({\bf a}^{-1}{y}, \tilde{x})$,
and then use the relation ${a}^{-1}{b}={h}^{-1}$ to also
relate the endpoint $y$ to its image in the fundamental domain,
$\tilde{\cal L} (\tilde{y},\tilde{x}) \equiv
{\cal L} ({\bf h}^{-1}\tilde{y}, \tilde{x})$.
With this operator which is restricted to the fundamental domain,
the global dynamics reduces to
\[
{\cal L}_{ba} (y,x) = D(h)_{ba}\tilde{{\cal L}} (\tilde{y},\tilde{x})
\,\, .
\]\noindent
While the global trajectory runs over the full space $M$, the restricted
trajectory is brought back into the fundamental domain $\tilde{M}$ any
time it crosses into adjoining tiles; the two trajectories
are related by the symmetry operation ${h}$ which maps the global
endpoint into its fundamental domain image.

Now the traces (\ref{det_tr}) required for the evaluation of the eigenvalues
of the transfer operator can be evaluated on the fundamental domain alone
\beq
\tr {\cal L} =
\int_{M}dx {\cal L} (x,x) =
  \int_{\tilde{M}} d\tilde{x}
\, \, \sum_{h}
\tr D(h) \, \,
{\cal L} ({\bf h}^{-1}\tilde{x},\tilde{x})
\label{M_to_fund}
\eeq
The fundamental domain integral
$  \int d\tilde{x} \, \, {\cal L} ({\bf h}^{-1}\tilde{x},\tilde{x}) $
picks up a contribution from every
global cycle (for which $h=e$), but it also picks up contributions from shorter
segments of global cycles. The permutation matrix $D(h)$ guarantees by the
identity $\tr D(h)=0$, ${h} \neq e$, that
only those repeats of the fundamental domain cycles $\tilde{p}$ that correspond
to complete global cycles $p$ contribute. Compare, for example, the
contributions of the $\overline{12}$ and $\overline{0}$
cycles of fig.~\ref{fg3_dsk}. $\tr D(h)\tilde{{\cal L}}$ does not get a
contribution from the $\overline{0}$ cycle, as the symmetry operation that
maps the first half of the $\overline{12}$ into the fundamental domain is
a reflection, and $\tr D(\sigma) =0$. In contrast, $\sigma^2=e$,
$\tr D(\sigma^2)=6$ insures that the repeat of the fundamental domain
fixed point $\tr (D(h)\tilde{{\cal L}})^2=6 t_0^2$, gives the correct
contribution to the global trace $\tr {\cal L}^2=3 \cdot 2 t_{12}$.

Let $p$ be the full orbit, $\tilde p$ the orbit in the fundamental
domain and $h_{\tilde p}$ an element of ${\cal H}_p$, the symmetry
group of $p$. Restricting the volume integrations to the infinitesimal
neighborhoods of the cycles $p$ and $\tilde p$, respectively,
and performing the standard resummations\rf{AACI}, we
obtain the identity
\beq
(1-t_p)^{m_p} =
        \det \left( 1 -  {D}({h}_{\tilde{p}}) t_{\tilde{p}}
        \right) \,\, ,
\label{eq1_rep}
\eeq
valid cycle by cycle in the Euler products (\ref{zeta__})
for $\det(1-{\cal L}) $.
Here ``det" refers to the
$\lbrack g \times g \rbrack$ matrix representation
$ 
{D}({h}_{\tilde{p}})$;
as we shall see, this determinant can be evaluated in terms of standard
characters, and no explicit representation of
${D}({h}_{\tilde{p}})$ is needed. Finally, if a cycle
${p}$ is invariant under the symmetry subgroup ${\cal H}_{{p}} \subseteq G$ of
order $h_{{p}}$, its weight can be written as a repetition of a fundamental
domain cycle
\beq
t_{{p}} =  t_{\tilde{p}}^{ h_{{p}} }
\label{t_power}
\eeq
computed on the irreducible segment that coresponds to a
fundamental domain cycle.
For example, in fig.~\ref{fg3_dsk} we see by inspection that
$t_{12}=t^2_0$ and $t_{123}=t^3_1$.

We conclude this section with a few comments about the role of symmetries
in actual extraction of cycles.
In the example at hand, the $N$-disk billiard systems, a fundamental domain is
a sliver of the $N$-disk configuration space delineated by a pair of
adjoining symmetry axes, with
the directions of the momenta indicated by arrows.
The flow may further be reduced to a return map on a Poincar\'e surface of
section, on which an appropriate transfer operator may be constructed.
While in principle any Poincar\'e surface of section will do,
a natural choice
in the present context are crossings of symmetry axes.

In actual numerical integrations only the last crossing of a symmetry line
needs to be determined (using for example the method of ref.~\cite{Henon}).
The cycle is run in global coordinates and the
group elements associated with the crossings of symmetry lines are
recorded; integration is terminated when the orbit closes in the fundamental
domain. Periodic orbits with non-trivial symmetry subgroups are particularly
easy to find since their points lie on crossings of symmetry
lines\rf{devo,rich}. A multi-point-shooting method combined with
Newton-Raphson iteration has proven very efficient\rf{PS} in practice.

\subsection{Boundary orbits}
\label{bound_o}

Before we can turn to a presentation of the factorizations of zeta functions
for the different symmetries we have to discuss a peculiar effect that
arises for orbits that run on a symmetry line that borders a fundamental
domain\rf{HW90,laur,BV}. In our 3-disk example, no such orbits are possible,
but they exist in other systems, such as
in the bounded region of the H\'enon-Heiles
potential and in 1-d maps. For the
symmetrical 4-disk billiard, there are in principle two kinds of such
orbits, one kind bouncing back and forth between two diagonally opposed disks
and the other kind moving along the other axis of reflection symmetry; the
latter exists for bounded systems only. While there are typically very few
boundary orbits, they tend to be among the shortest orbits, and their neglect
can seriously degrade the convergence of cycle expansions, as those are
dominated by the shortest cycles.

While such orbits are invariant under some symmetry operations,
their neighborhoods are not. This affects the stability
matrix ${\bf J}_p$ of the
linearization perpendicular to the orbit and thus the eigenvalues. 
Typically, 
{\em e.g.} if the symmetry is a reflection, some eigenvalues of ${\bf J}_p$
change sign. This means that instead of a weight $1/\det({\bf 1}-{\bf J}_p)$
as for a regular orbit, boundary cycles also pick up contributions of form
$1/\det({\bf 1}-{\bf h}{\bf J}_p)$, 
where ${\bf h}$ is a symmetry operation that
leaves the orbit pointwise invariant; see for example sect.~\ref{Reflecti}.

Consequences for the zeta function factorizations are that sometimes
a boundary orbit does not contribute. A derivation of a dynamical zeta function
(\ref{zeta__}) from a determinant like (\ref{det(1-L)})
usually starts with an expansion
of the determinants of the Jacobian. The leading order terms just contain
the product of the expanding eigenvalues and lead to
the zeta function (\ref{zeta__}).
Next to leading order terms contain products of expanding and contracting
eigenvalues and are sensitive to their signs. Clearly, the weights
$t_p$ in the zeta functions will then be affected by reflections in the
Poincar\'e surface of section perpendicular to the orbit.
In all our applications it was possible to implement these effects
by the following simple prescription.

If an orbit is invariant under a little group
${\cal H}_p = \{e, b_2,\dots,b_h\}$, then the corresponding group element in
(\ref{eq1_rep}) will be replaced by a projector. If the weights are insensitive
to the signs of the eigenvalues, then this projector is
\beq
{ g}_p = \frac{1}{h} \sum_{i=1}^h { b}_i \,.
\label{L_bound}
\eeq
In the cases that we have considered,
the change of sign may be taken into account by defining a sign function
$\epsilon_p({ g}) = \pm 1$, with the ``-" sign if the symmetry
element ${g}$ flips the neigborhood. Then (\ref{L_bound}) is replaced by
\beq
{ g}_p = \frac{1}{h} \sum_{i=1}^h \epsilon({ b}_i) \, { b}_i \,.
\label{L_bound_ext}
\eeq
The resulting zeta functions agree with the ones given by Lauritzen\rf{laur}.
We illustrate the above in sect.~\ref{Reflecti}
by working out the full factorization for the 1-dimensional
reflection symmetric maps.

\section{Factorizations of zeta functions}
\label{Factor}

In the above we have shown that a discrete symmetry induces degeneracies among
periodic orbits and decomposes periodic orbits into repetitions of irreducible
segments; this reduction to a fundamental domain furthermore leads to a
convenient symbolic dynamics compatible with the symmetry, and, most
importantly, to a factorization of zeta functions. This we now develop, first
in a general setting and then for specific examples.

\subsection{Factorizations of dynamical zeta functions}
According to (\ref{eq1_rep}) and (\ref{t_power}),
the contribution of a degenerate class
of global cycles (cycle $p$ with multiplicity $m_p=g/h_p$) to a zeta function
is given by the corresponding fundamental domain  cycle $\tilde{p}$:
\beq
(1-t_{\tilde{p}}^{h_p})^{g/h_p}
    =\det \left(1- D(h_{\tilde p}) t_{\tilde p} \right)
\label{cyc_fac}
\eeq
Let $D(h) = \bigoplus_{\alpha} d_{\alpha} D_{\alpha}(h)$
be the decomposition of the
matrix representation $D(h)$ into the $d_\alpha$ dimensional irreducible
representations $\alpha$ of a finite group $G$. Such decompositions are
block-diagonal, so the corresponding contribution to the Euler product
(\ref{det(1-L)}) factorizes as
\beq
\det(1- D(h) t ) =
  \prod_{\alpha} \det(1-D_{\alpha}(h) t )^{d_\alpha}
\,\, ,
\label{factzet}
\eeq
where now the product extends over all distinct $d_\alpha$-dimensional
irreducible representations, each contributing $d_\alpha$ times.
For the cycle expansion purposes, it has been convenient to emphasize that the
group-theoretic factorization can be effected cycle by cycle, as in
(\ref{cyc_fac}); but from the transfer operator point of view, the key
observation is that the symmetry reduces the transfer operator to a block
diagonal form; this block diagonalization implies that the
zeta functions
(\ref{zeta__}) factorize as
\beq
\frac{1}{\zeta} = \prod_\alpha \frac{1}{\zeta_\alpha^{d_\alpha}}
\,\, ,\quad \quad
\frac{1}{\zeta_\alpha} = \prod_{\tilde{p}} \det
        \left( 1 -  D_\alpha( h_{\tilde{p}}) t_{\tilde{p}}
        \right) \,\, .
\label{zet_fac}
\eeq

Determinants of $d$-dimensional irreducible representations can be
evaluated using the expansion of determinants in terms of traces,
\bea
\det (1+M) &=& 1 + \tr M + {1 \over 2}\left( (\tr M)^2 - \tr M^2 \right)
\ceq
 + {1 \over 6} \left( (\tr M)^3 - 3 \, (\tr M) (\tr M^2) + 2 \,\tr M^3\right)
\ceq
+ \cdots + {1 \over d!} \left( (\tr M)^d - \cdots \right)
\,\, ,
\label{detM}
\eea
and each factor in (\ref{factzet}) can be evaluated by looking up the
characters $\chi_{\alpha}(h)= \tr D_{\alpha}(h)$ in standard
tables\rf{hamer}. In terms of characters, we have for the 1-dimensional
representations
\[
 \det(1-D_{\alpha}(h) t) = 1 - \chi_{\alpha}(h) t
\,\, ,
\]\noindent
for the 2-dimensional representations
\[
\det(1-D_{\alpha}(h) t) = 1 - \chi_{\alpha}(h) t +
{1 \over 2} \left(
\chi_{\alpha}(h)^2 - \chi_{\alpha}(h^2)\right) t^2 ,
\]\noindent
and so forth. These expressions can sometimes be simplified further using
standard group-theoretical methods. For example, the $ {1 \over 2}
\left( (\tr M)^2 - \tr M^2 \right) $ term in (\ref{detM}) is the trace
of the antisymmetric
part of the $M \times M$ Kronecker product; if $\alpha$ is a 2-dimensional
representation, this is the $A_2$ antisymmetric representation, so
\beq
\mbox{\rm 2-dim:} \quad
\det(1-D_{\alpha}(h) t) = 1 - \chi_{\alpha}(h) t +
\chi_{A_2}(h) t^2 .
\label{s2d}
\eeq

In the fully symmetric subspace $\tr D_{A_1}(h)=1$ for all orbits;
hence a straightforward fundamental domain computation (with no
group theory weights) always yields a part of the full spectrum.
In practice this is the most interesting subspectrum, as it
contains the leading eigenvalue of the transfer operator.

\subsection{Factorizations of functional determinants}
\label{Factfunc}

Factorization of the full functional determinant (\ref{det_tr})
proceeds in essentially the same manner as
the factorization of dynamical zeta functions outlined above.
By (\ref{L_alpha}) and (\ref{M_to_fund})
the trace of the transfer operator ${\cal L}$
splits into the sum of inequivalent irreducible subspace
contributions
$ \sum_\alpha \tr {\cal L}_\alpha$, with
\[
\tr {\cal L}_\alpha = d_\alpha
\sum_{h \in G} \chi_\alpha (h) \int_{\tilde{M}} d\tilde{x}
{\cal L} ({\bf h}^{-1} \tilde{x},\tilde{x})
\, .
\]\noindent
This leads by standard manipulations\rf{AACI,CE} to
the factorization of (\ref{det(1-L)}) into
\beq
Z(z) = \prod_\alpha Z_\alpha (z)^{d_\alpha}
\,\, , \quad \quad
Z_\alpha (z) =
{\rm exp}  \left( - {
         \sum_{\tilde{p}} \sum_{r=1}^\infty {1 \over r}
 {\chi_\alpha (h_{\tilde{p}}^r)  z^{n_{\tilde{p}} r}
 \over  | \det \left( {\bf 1}-{\bf {\tilde{J}}}_{\tilde{p}}^{r} \right) | }
         } \right)
\,\,  ,
\label{det_fac}
\eeq
where
${\bf \tilde{J}}_{\tilde{p}} =
{\bf h}_{\tilde{p}} {\bf J}_{\tilde{p}} $
is the fundamental domain Jacobian.
Boundary orbits
require special treatment, discussed in sect.~\ref{bound_o},
with examples given in the next section as well as in
sect.~\ref{exampl}.

\vskip 12pt

The factorizations (\ref{zet_fac}), (\ref{det_fac})
are the main result of this paper. We now proceed to
examplify it by a few cases of physical interest.
Additional examples
(factorization and cycle expansions for the cyclic symmetry groups
$C_3$ and $C_{N}$) are given in ref.~\cite{russ}.

\subsection{Reflection symmetric 1-d maps: $C_2$ factorization}
\label{Reflecti}

Consider $f$, a map on the interval
with reflection symmetry $f(-x) = - f(x)$.
Denote the reflection operation by ${\bf C}x = - x$.
The symmetry of the map implies that if $\{x_n\}$ is a
trajectory, than also $\{{\bf C} x_n\}$ is a trajectory because
$
{\bf C} x_{n+1} = {\bf C} f(x_n) = f({\bf C} x_n) \, .
$
The dynamics can be restricted to a fundamental domain,
in this case to one half of the original interval;
every time a
trajectory leaves this interval, it can be mapped back using ${\bf C}$.

To compute the traces of the symmetrization and  antisymmetrization
projection operators (\ref{C_proj}), we have to distinguish
three kinds of cycles\rf{Zakopane}: asymmetric cycles $\asym $,
symmetric cycles $\sym$ built by repeats of
irreducible segments $\symf$, and boundary cycles $b$.
The Fredholm determinant can be formally written as the
product over the three kinds of cycles:
$
\det(1- \Lop) =
\det(1- \Lop)_\asym
\det(1- \Lop)_\symf
\det(1- \Lop)_b
$.
\\

\noindent
{\bf Asymmetric cycles:}
A periodic orbits is not symmetric if
$\{x_\asym\} \cap \{{\bf C} x_\asym\} = \emptyset$, where
$\{x_\asym\}$ is the set of periodic points belonging to the cycle $\asym$.
Thus ${\bf C} $ generates a second orbit
with the same number of points and the same stability properties.
Both orbits give the same contribution to the first term and
no contribution to the second term in (\ref{C_proj});
as they are degenerate, the prefactor $1/2$ cancels.
Resumming as in the derivation of (\ref{fredh}), we find that
asymmetric orbits yield the same contribution to the symmetric
and the antisymmetric subspaces:
\[
\det(1- \Lop_{\pm})_\asym =
   \prod_{\asym}\  \prod_{k=0}^\infty
          \left(1 -  { t_\asym \over { \Lambda_\asym^{k}} } \right)
   \, , \quad
t_\asym = { z^{n_\asym} \over {|\Lambda_\asym|}} \, .
\]\noindent

\vskip 12pt
\noindent
{\bf Symmetric cycles:}
A cycle $\sym$ is reflection symmetric if
operating with ${\bf C} $ on the set of
cycle points reproduces the set. The period of a symmetric cycle
is always even ($\nsym = 2 \nsymf$) and the mirror image of the
$x_\sym$ cycle point is reached by traversing the irreducible
segment $\symf$ of length $\nsymf$, $f^{\nsymf}(x_\sym) = {\bf C}  x_\sym $.
$\delta(x-f^n(x))$ picks up $2 \nsymf$ contributions
for every even traversal, $n=r \nsymf $, $r$ even,  and
$\delta(x+f^n(x))$ for every odd traversal, $n=r \nsymf $, $r$ odd.
Absorb the group-theoretic prefactor in the stability eigenvalue
by defining $\Lambda_\symf = - Df^{\nsymf}(x_\sym)$,
where $Df^{\nsymf}(x_\sym)$ is the stability
computed for a segment of
length $\nsymf$. Restricting the integration to the infinitesimal
neighborhood of the
$\sym$ cycle, we obtain the contribution to $ tr{\cal L}_{\pm}^n$:
\bea
z^n \tr {\cal L}^n_{\pm}
   & \rightarrow & \int_{V_\sym} dx\,
   z^n \, {1\over 2} \,  \left(
\delta(x-f^n(x)) \pm  \delta(x+f^n(x)) \right)
                                                \nonumber\\
& = & \nsymf \left(
   \sum^{\rm even}_{r=2}
     \delta_{n,r \nsymf} { {t^{r}_\symf} \over {1-1/\Lambda_\symf^{r}}}
   \pm
   \sum^{\rm odd}_{r=1}
     \delta_{n,r \nsymf} { {t^{r}_\symf} \over {1-1/\Lambda_\symf^{r}}}
         \right)
                                                \nonumber\\
& = & \nsymf \sum^{\infty}_{r=1}
 \delta_{n,r \nsymf} { {(\pm t_\symf)^{r}} \over {1-1/\Lambda_\symf^r} }
\, .
                                                \nonumber
\eea
Substituting all  symmetric cycles $\sym$ into $\det(1-\Lop_{\pm})$
and resumming as in (\ref{fredh}), we obtain:
\[
\det(1-\Lop_{\pm})_\symf=
   \prod_{\symf}\  \prod_{k=0}^\infty
       \left(1 \mp { t_\symf \over { \Lambda_\symf^k } } \right)
\]

\vskip 12pt
\noindent
{\bf Boundary cycles:}
In the example at hand
there is only one cycle which is neither symmetric
nor antisymmetric, but lies on the boundary of the fundamental
domain, the fixed point
at the origin. Such a cycle contributes simultaneously to both
$\delta(x-f^n(x))$ and
$\delta(x+f^n(x))$:
\bea
z^n \tr {\cal L}^n_{\pm}
   & \rightarrow & \int_{V_b} dx  \,
   z^n  \, {1\over 2} \, \left(
\delta(x-f^n(x)) \pm  \delta(x+f^n(x)) \right)
                                                \nonumber\\
& = & \sum^{\infty}_{r=1}
     \delta_{n,r} \, t^r_b \,
     {1 \over 2} \,
        \left(
     {1 \over {1-1/\Lambda_b^{r}} }
                \pm
     {1 \over {1+1/\Lambda_b^{r}} }
         \right)
                                                \nonumber\\
z^n \, \tr{\cal L}^n_{+}
& \rightarrow & \sum^{\infty}_{r=1}
 \delta_{n,r} { {t_b^r} \over {1-1/\Lambda_b^{2r}} }
\, ; \quad \quad
z^n \, \tr{\cal L}^n_{-}
\rightarrow  \sum^{\infty}_{r=1}
 \delta_{n,r} { 1 \over \Lambda_b^{r} }
              { {t_b^r} \over {1-1/\Lambda_b^{2r}} }
\, .
                                                \nonumber
\eea
Boundary orbit contributions to the factorized Fredholm
determinants follow by resummation:
\[
\det(1-\Lop_{+})_b=
    \prod_{k=0}^\infty
       \left(1 - { t_b \over { \Lambda_b^{2k}} } \right)
\, , \quad \quad
\det(1-\Lop_{-})_b=
\prod_{k=0}^\infty
       \left(1 - { t_b \over { \Lambda_b^{2k+1}} } \right)
\]\noindent
Only even derivatives contribute
to the symmetric subspace (and odd to the antisymmetric
subspace) because the orbit lies on the boundary.
The symmetry reduced Fredholm determinants follows by collecting
the above results:
\[
Z_{+}(z) =
   \prod_{\asym}\  \prod_{k=0}^\infty
          \left(1 -  { t_\asym \over { \Lambda_\asym^{k}} } \right)
   \prod_{\symf}\  \prod_{k=0}^\infty
       \left(1 - { t_\symf \over { \Lambda_\symf^k } } \right)
                 \prod_{k=0}^\infty
       \left(1 - { t_b \over {\Lambda_b^{2k}} } \right)
\]
\beq
Z_{-}(z) =
   \prod_{\asym}\  \prod_{k=0}^\infty
          \left(1 -  { t_\asym \over { \Lambda_\asym^{k}} } \right)
   \prod_{\symf}\  \prod_{k=0}^\infty
       \left(1 + { t_\symf \over { \Lambda_\symf^k } } \right)
                 \prod_{k=0}^\infty
       \left(1 - { t_b \over { \Lambda_b^{2k+1}} } \right)
\label{omeg_a}
\eeq
As reflection symmetry is essentially the only
discrete symmetry that a map of the interval can have, this
example completes the group-theoretic factorization of
determinants and zeta functions for 1-dimensional maps.
A specific example is worked out in Ref.~\cite{Zakopane}

\section{Examples of symmetry induced factorizations}
\label{exampl}

We conclude with several explicit examples of group theory
factorizations of cycle expansions of dynamical zeta functions.
These expansions are a prerequisite for applications of periodic orbit theory
to the evaluation of classical and quantal spectra; in particular, they
were used in the calculations of refs.~\cite{eck,CES}.

\subsection{$C_2$ factorizations}

As the simplest example of implementing the above scheme
consider the $C_2$ symmetry which arises, for example, in the Lorenz
system\rf{GO}, in the 3-dimensional anisotropic Kepler
problem\rf{gutz_sym,TW,CC92} or in the cycle expansions treatments of the Ising
model\rf{ronnie}. For our purposes, all that we need to know here  is that each
orbit or configuration is uniquely labelled by an infinite string $\{s_i\}$,
$s_i=+,-$ and that the dynamics is invariant under the
$ + \leftrightarrow - $ interchange, {\em i.e.}, it is $C_2$ symmetric.
In the Lorenz system, the labels $+$ and $-$ stand for the left or the right
lobe of the attractor and the symmetry is
a rotation by $\pi$ around the z-axis.
Similarly, the Ising Hamiltonian
(in the absence of an external magnetic field)
is invariant under spin flip.
The $C_2$ symmetry cycles separate into two classes, the
self-dual configurations
$+-$, $++--$, $+++---$, $+--+-++-$, $\cdots$, with multiplicity
$m_p=1$, and the asymmetric configurations $+$, $-$, $++-$, $--+$, $\cdots$,
with multiplicity $m_p=2$. For example, as there is no absolute distinction
between the ``up" and the ``down" spins, or the ``left" or the ``right" lobe,
$t_+=t_-$,  $t_{++-}=t_{+--}$, and so on.

The symmetry reduced labelling $\rho_i \in \{0,1\}$
is related to the standard $s_i \in \{+,-\}$ Ising spin labelling by
\bea
\mbox{If} \quad s_{i} & = & s_{i-1} \quad \mbox{then} \quad \rho_i=1
     \continue
\mbox{If} \quad s_{i} & \neq & s_{i-1} \quad \mbox{then} \quad \rho_i=0
\eea
For example,
$\overline{+}=\cdots ++++ \cdots$ maps into $\cdots 111 \cdots = \overline{1}$
(and so does $\overline{-}$),
$\overline{-+}=\cdots -+-+ \cdots$ maps into $\cdots 000 \cdots = \overline{0}$,
$\overline{-++-}=\cdots --++--++ \cdots$
maps into $\cdots 0101 \cdots = \overline{01}$, and so forth.
A list of such reductions is given in table 4.

Depending on the maximal symmetry group ${\cal H}_p$ that
leaves an orbit $p$ invariant ({\em cf}. sects.~\ref{degene}, \ref{Dynami}),
the contributions to the zeta function factor as
\bea
                     &  & ~~~~A_1~~~~~~A_2
                \continue
{\cal H}_{{p}} = \{e\}: \quad
( 1 - t_{\tilde p} )^2 &  = & (1 - t_{\tilde p})(1 - t_{\tilde p})
                \continue
{\cal H}_{{p}} = \{e,\sigma\}: \quad
( 1 - t^2_{\tilde p} ) & = & (1 - t_{\tilde p}) (1 + t_{\tilde p})
\,\, ,
\label{symm_isin}
\eea
For example:
\bea
{\cal H}_{{++-}} = \{e\}: \quad
( 1 - t_{++-} )^2 &  = & (1 - t_{001})(1 - t_{001})
                \continue
{\cal H}_{{+-}} = \{e,\sigma\}: \quad
( 1 - t_{+-} )~~ & = & ~(1 - t_{0})~~(1 + t_{0}), \quad t_{+-}=t_{0}^2
                                        \nonumber
\eea
This yields two binary expansions. The $A_1$ subspace zeta function is given
by the standard binary expansion (\ref{curvbin}). The antisymmetric $A_2$
subspace zeta function $\zeta_{A_2}$
differs from $\zeta_{A_1}$ only by a minus sign for
cycles with an odd number of $0$'s:
\bea
1/\zeta_{A_2} & = & (1+t_{0})(1-t_{1})(1+t_{10})(1-t_{100})(1+t_{101})
            (1+t_{1000}) \ceq
            (1-t_{1001})(1+t_{1011})
            (1-t_{10000})(1+t_{10001}) \ceq (1+t_{10010})
            (1-t_{10011})(1-t_{10101})(1+t_{10111})
            \dots \continue
 &= &  1 + t_0 - t_1 + (t_{10} - t_1 t_0)
 - (t_{100}- t_{10} t_0)
 + (t_{101}- t_{10} t_1) \ceq
 - (t_{1001}  - t_1 t_{001} - t_{101} t_0 + t_{10}t_0 t_1) - \dots
\dots
\label{zetaA2}
\eea
Note that the group theory factors do not destroy the
curvature corrections (the cycles and pseudo cycles are
still arranged into shadowing combinations).

If the system under consideration has a boundary orbit
({\em cf.} sect.~\ref{bound_o}) with
group-theoretic factor ${\bf h}_p=({\bf e+\bf \sigma})/2$,
the boundary orbit does not contribute to the antisymmetric
subspace
\bea
& & \quad A_1 \quad \quad A_2   \nonumber\\
\mbox{boundary:} \quad (1-t_{p}) & = &
(1-t_{\tilde{p}})(1-0 t_{\tilde{p}})
\eea
This is the $1/\zeta$ part of the boundary orbit factorization of
sect.~\ref{Reflecti}.

\subsection{3-disc pinball: $C_{3v}$ factorization}
\label{C3v_fact}

The next example, the $C_{3v}$ symmetry, can be worked out by a glance at
fig.~\ref{fg3_dsk}a.
For the symmetric $3$-disk pinball the fundamental domain is bounded by
a disk segment and the two adjacent sections of the symmetry axes that act as
mirrors (see fig.~\ref{fg3_dsk}b).
The three symmetry axes divide the space into six copies
of the fundamental domain. Any trajectory on the full space can be pieced
together from bounces in the fundamental domain, with symmetry axes replaced
by flat mirror reflections. The binary $\{0, 1\}$ reduction of
the ternary three disk $\{1,2,3\}$ labels has a simple geometric interpretation:
a collision of type $0$ reflects the projectile to the disk it
comes from (back--scatter), whereas after a collision of type $1$
projectile continues to the third disk.
For example,
$\overline{23}=\cdots 232323 \cdots$ maps into
$\cdots 000 \cdots =  \overline{0}$
(and so do $\overline{12}$ and  $\overline{13}$),
$\overline{123}=\cdots 12312 \cdots$ maps into $\cdots 111 \cdots =
\overline{1}$ (and so does $\overline{132}$), and so forth. A list of such
reductions for short cycles is given in table~5.

$C_{3v}$ has two one-dimensional irreducible representations, symmetric
and antisymmetric under reflections, denoted $A_1$ and $A_2$, and 
a pair of 
degenerate two-dimensional representations of mixed symmetry, denoted $E$.
The contribution of an orbit with symmetry $g$ to the $1/\zeta$ Euler product
(\ref{factzet}) factorizes according to
\bea
\det(1-D(h)t) &=& \left(1 - \chi_{A_1}(h) t \right)
\left(1 - \chi_{A_2}(h) t \right)
\left(1 - \chi_{E}(h) t + \chi_{A_2}(h) t^2 \right)^2
\label{fact3d}
\eea
with the three factors contributing to the $C_{3v}$ irreducible
representations $A_1$, $A_2$ and $E$, respectively, and
the 3-disk zeta function
factorizes into   $ \zeta =  \zeta_{A_1} \zeta_{A_2} \zeta_E^2 $.
Substituting the $C_{3v}$ characters\rf{hamer}
\vskip .3cm
\begin{center}
\begin{tabular}{|c|crr|}
\hline
$C_{3v}$     & $A_1$& $A_2$& $E$ \\
\hline
$ e   $      &   1  &   1  &  2  \\
$ C_3,C_3^2$ &   1  &   1  & $-1$  \\
$ \sigma_v$  &   1  &  $-1$  &  0  \\
\hline
\end{tabular}
\end{center}
\vskip .3cm
\noindent
into (\ref{fact3d}), we obtain for the three classes of
possible orbit symmetries
(indicated in the first column)
\bea
  {\bf h}_{\tilde{p}}~~~~~~~~~~~~
&\ & \quad A_1 \quad \quad A_2 \quad \quad E \continue
e: \quad
( 1 - t_{\tilde p} )^6 &  = & (1 - t_{\tilde p})(1 - t_{\tilde p})
  (1 - 2 t_{\tilde p} + t_{\tilde p}^{2})^2 \continue
C_3,C_3^2: \quad
( 1 - t^3_{\tilde p} )^2 & = & (1 - t_{\tilde p}) (1 - t_{\tilde p})
                (1 + ~t_{\tilde p} + t_{\tilde p}^{2})^2 \continue
\sigma_v: \quad
( 1 - t^2_{\tilde p} )^3 & = & (1 - t_{\tilde p}) (1 + t_{\tilde p})
                (1 + 0 t_{\tilde p} - t_{\tilde p}^{2})^2 .
\label{symm}
\eea
where $\sigma_v$ stands for any one of the three reflections.

The Euler product (\ref{zeta__}) on each irreducible subspace follows from
the factorization (\ref{symm}). On the symmetric $A_1$ subspace the
$\zeta_{A_1}$ is given by the standard binary curvature expansion
(\ref{zetabin}). The antisymmetric $A_2$ subspace $\zeta_{A_2}$ differs
from $\zeta_{A_1}$ only by a minus sign for cycles with an odd number of
$0$'s, and is given in (\ref{zetaA2}). For the mixed-symmetry subspace $E$
the curvature expansion is given by
\bea
1/\zeta_E & = & (1+zt_{1}+z^2t_{1}^2)(1-z^2t_{0}^2)
                (1+z^3t_{100}+z^6t_{100}^2)  (1-z^4t_{10}^2) \ceq
                (1+z^4t_{1001}+z^8t_{1001}^2)
                (1+z^5t_{10000}+z^{10}t_{10000}^2) \ceq
                (1+z^5t_{10101}+z^{10}t_{10101}^2) 
                (1-z^5t_{10011})^2
                 \dots  \continue
        & = & 1+z t_{1}+z^2(t_{1}^2 -t_{0}^2)
               +z^3 (t_{001}-t_{1} t_{0}^2) \ceq
        + z^4 \left[ t_{0011}+(t_{001} -t_{1} t_{0}^2  ) t_{1}
          -t_{01}^2 \right]    \ceq
         + z^5 \left[t_{00001}+t_{01011}-2 t_{00111}
         +(t_{0011} - t_{01}^2) t_{1}
  +( t_{1}^2-t_{0}^2) t_{100} \right]
  + \cdots
\label{3dzetam}
\eea
We have reinserted the powers of $z$ in order to group together cycles
and pseudo-cycles of the same length. Note that the factorized
cycle expansions retain the curvature form; long cycles are still
shadowed by (somewhat less obvious) combinations of pseudocycles.

Refering back to the topological polynomial (\ref{3disk_top}) obtained by
setting $t_p=1$, we see that its factorization is a
consequence of the $C_{3v}$ factorization of the $\zeta$ function:
\[
1/\zeta_{A_1} = 1- 2 z \,\, , \quad
1/\zeta_{A_2}  = 1 \,\, , \quad
1/\zeta_{E} = 1 +  z \,\, ,
\]\noindent
as obtained from (\ref{zetabin}), (\ref{zetaA2}) and (\ref{3dzetam})
for $t_p = 1$.

An example of a system with $C_{3v}$ symmetry is provided by
the motion of a particle in the H\'enon-Heiles potential\rf{HH}
\[
V(r,\theta ) = {1 \over 2} r^2 + {1\over 3} r^3 \sin(3\theta )
\,\, .
\]\noindent
Our coding is not directly applicable to this system because of the existence
of elliptic islands and because the three orbits that run along the symmetry
axis cannot be labelled in our code. However, since these orbits run along the
boundary of the fundamental domain, they require the special treatment\rf{laur}
discussed in sect.~\ref{bound_o}.

Their symmetry is $K=\{ {\bf e},\sigma \}$, so according to (\ref{L_bound}),
they pick up the group-theoretic factor ${\bf h}_p=({\bf e+\sigma})/2$.
If there is no sign change in $t_p$, then evaluation of
$\det (1 - \frac{{\bf e + \sigma}}{2} t_{\tilde{p}})$ yields
\bea
&\ & \quad A_1 \quad \quad A_2 \quad \quad E \continue
\mbox{boundary:} \quad (1-t_{p})^3 & =&
(1-t_{\tilde{p}})(1-0 t_{\tilde{p}}) (1-t_{\tilde{p}})^2
\, , \quad \quad t_p=t_{\tilde{p}}
\, .
\label{bounda}
\eea
However, if the cycle weight changes sign under reflection,
$t_{\sigma \tilde{p}}= -t_{\tilde{p}} $,
the boundary orbit does not contribute to
the subspace
symmetric under reflection across the orbit;
\bea
&\ & \quad A_1 \quad \quad A_2 \quad \quad E \continue
\mbox{boundary:} \quad (1-t_{p})^3 & = &
(1-0t_{\tilde{p}})(1-t_{\tilde{p}}) (1-t_{\tilde{p}})^2
\, , \quad \quad t_p=t_{\tilde{p}}
\, .
\label{bounda_1}
\eea

\subsection{$C_{4v}$ factorization}
\label{C_4v_inva}

If an $N$-disk arrangement has $C_N$ symmetry, and the disk visitation
sequence is given by disk labels $\{ \epsilon_1 \epsilon_2 \epsilon_3 \dots\}$,
only the relative increments
$\rho_i = \epsilon_{i+1}-\epsilon_{i} \,\, {\rm mod\ } N$ matter.
Symmetries under reflections across axes increase the group to $C_{Nv}$ and
add relations between symbols: $\{\epsilon_i\}$ and $\{N - \epsilon_i\}$
differ only by a reflection\rf{tdisk,grass}. As a consequence of this
reflection increments become decrements until the next reflection and
vice versa. Consider four equal disks placed on the vertices
of a square (fig.\ref{fgC4v}). The symmetry group consists
of the identity $\bf e$, the two reflections $\sigma_x$, $\sigma_y$
across $x$, $y$ axes, the two diagonal reflections
$\sigma_{13}$, $\sigma_{24}$,
and the three rotations $C_4$, $C_2$ and $C_4^3$ by
angles $\pi/2$, $\pi$ and $3\pi/2$.
We start by exploiting the $C_4$ subgroup symmetry in order to
replace the absolute labels
$\epsilon_i \in \{ 1,2,3,4\}$ by relative increments $\rho_i \in \{ 1,2,3\}$.
By reflection across diagonals, an
increment by $3$ is equivalent to an increment by $1$ and a reflection;
this new symbol will be called $\underline{1}$.
Our convention will be to first perform
the increment and then to change the orientation due to the reflection.
As an example, consider the fundamental domain cycle
$112$. Taking the disk 1 $\rightarrow$
disk 2 segment as the starting segment, this symbol string
is mapped into the disk visitation sequence
$1_{+1} 2_{+1} 3_{+2}1 \dots= \overline{123}$, where the subscript
indicates the increments (or decrements) between neighboring symbols;
the period of the cycle $\overline{112}$
is thus 3 in both the fundamental domain and
the full space. Similarly, the cycle
$\overline{\underline{1}12}$ will be mapped into
$1_{+1} 2_{-1} 1_{-2} 3_{-1} 2_{+1} 3_{+2}1= \overline{121323}$
(note that the fundamental domain symbol $\underline{1}$
corresponds to a flip in orientation
after the second and fifth symbols); this time the period in the full space
is twice that of the fundamental domain. In particular, the
fundamental domain fixed points correspond to the following
4-disk cycles:
\vskip 12pt
\begin{center}
\begin{tabular}{lcr}
4-disk            &                    & reduced \\
$ {12} $ & $ \leftrightarrow $ & $ {\underline{1}} $ \\
$ {1234} $ & $ \leftrightarrow $ & $ {1} $ \\
$ {13} $ & $ \leftrightarrow $ & $ {2} $ \\
\end{tabular}
\end{center}
\vskip 12pt
Conversions for all periodic orbits of reduced symbol period less than 5 are
listed in table~6.

While there is a variety of labelling conventions\rf{freddy,grass,EW1} for the
reduced $C_{4v}$ dynamics, we prefer the one introduced here because of its
close relation to the group-theoretic structure of the dynamics: the global
4-disk trajectory can be generated by mapping the fundamental domain
trajectories onto the
full 4-disk space by the accumulated product of the $C_{4v}$ group elements
$g_1=C$, $g_2=C^2$, $g_{\underline{1}}=\sigma_{diag} C= \sigma_{axis}$,
where $C$ is a rotation by $\pi/2$. In the $\overline{\underline{1}12}$
example worked out above, this yields
$g_{\underline{1}12} = g_{2} g_{1} g_{\underline{1}}
= C^2  C \sigma_{axis} = \sigma_{diag}$, listed in the last column of
table~6.
Our convetion is to multiply group elements in the reverse order with
respct to the symbol sequence.
We need these group elements for our next step, the zeta function
factorizations.

The $C_{4v}$ group has four one-dimensional representations, either symmetric
($A_1$) or antisymmetric ($A_2$) under both types of reflections, or symmetric
under one and antisymmetric under the other ($B_1$, $B_2$), and a
degenerate pair of
two-dimensional representations $E$. Substituting the $C_{4v}$ characters
\vskip .3cm
\begin{center}
\begin{tabular}{|c|crrrr|}
\hline
$C_{4v}$       & $A_1$& $A_2$& $B_1$& $B_2$& $E$ \\
\hline
$  e  $        &   1  &   1  &  1   &  1   &  2  \\
$C_2  $        &   1  &   1  &  1   &  1   & -2  \\
$C_4,C_4^3  $  &   1  &   1  & -1   & -1   &  0  \\
$\sigma_{axes}$&   1  &  -1  &  1   & -1   &  0  \\
$\sigma_{diag}$&   1  &  -1  & -1   &  1   &  0  \\
\hline
\end{tabular}
\end{center}
\vskip .3cm
\noindent
into (\ref{zet_fac}) we obtain:
\vskip 12pt
\begin{tabular}{rlcccccc}

$h_{\tilde p}$ &  & &  $A_1$  &  $A_2$  &  $B_1$  &  $B_2$  &  $E$  \\
$e$:
& $(1-t_{\tilde p} )^8$  &=&$(1-t_{\tilde p})$ & $(1-t_{\tilde p})$ &
                        $(1-t_{\tilde p})$ &$(1-t_{\tilde p})$&$ (1-t_{\tilde
 p})^4 $ \\
$C_2$:
& $(1-t_{\tilde p}^2 )^4$ &=&  $(1-t_{\tilde p})$ & $(1-t_{\tilde p})$ &
                        $(1-t_{\tilde p})$ &$(1-t_{\tilde p})$&$ (1+t_{\tilde
 p})^4 $ \\
$C_4,C_4^3$:
& $(1-t_{\tilde p}^4 )^2$ &=&  $(1-t_{\tilde p})$ & $(1-t_{\tilde p})$ &
                        $(1+t_{\tilde p})$ &$(1+t_{\tilde p})$&$ (1+t_{\tilde
 p}^2)^2 $ \\
$\sigma_{axes}$:
& $(1-t_{\tilde p}^2 )^4$&=& $(1-t_{\tilde p})$ & $(1+t_{\tilde p})$ &
                        $(1-t_{\tilde p})$ &$(1+t_{\tilde p})$&$ (1-t_{\tilde
 p}^2)^2 $ \\
$\sigma_{diag}$:
& $(1-t_{\tilde p}^2 )^4$&=& $(1-t_{\tilde p})$ & $(1+t_{\tilde p})$ &
                        $(1+t_{\tilde p})$ &$(1-t_{\tilde p})$&$ (1-t_{\tilde
 p}^2)^2 $ \\
\end{tabular}
\vskip 12pt
\noindent
The possible irreducible segment
group elements ${\bf h}_{\tilde{p}}$ are listed in the first
column; $\sigma_{axes}$ denotes a reflection across either the x-axis or the
y-axis, and $\sigma_{diag}$ denotes a reflection across a diagonal
(see fig.~\ref{fgC4v}). In addition, degenerate pairs of boundary orbits can
run along the symmetry lines in the full space, with the fundamental domain
group theory weights
${\bf h}_p=(C_2+\sigma_x)/2$ (axes) and
${\bf h}_p=(C_2+\sigma_{13})/2$ (diagonals) respectively:
\bea
  & & \quad A_1   \quad \quad  A_2  \quad \quad  B_1
      \quad \quad  B_2  \quad \quad  E  \continue
\mbox{axes:}\quad  (1-t_{\tilde{p}}^2)^2 & = &
(1-t_{\tilde{p}})(1-0 t_{\tilde{p}})
(1-t_{\tilde{p}})(1-0 t_{\tilde{p}}) (1+t_{\tilde{p}})^2
\continue
\mbox{diagonals:}\quad  (1-t_{\tilde{p}}^2)^2 & = &
(1-t_{\tilde{p}})(1-0 t_{\tilde{p}})
(1-0 t_{\tilde{p}})(1- t_{\tilde{p}}) (1+t_{\tilde{p}})^2
\label{bounda4}
\eea
(we have assumed that $t_{\tilde{p}}$ does not change sign under
reflections across symmetry axes).
For the 4-disk arrangement considered here
only the diagonal orbits $\overline{13}$,
$\overline{24}$ occur; they correspond to the
$\overline{2}$ fixed point in the fundamental domain.

The $A_1$ subspace in $C_{4v}$ cycle expansion
is given by
\bea
1/\zeta_{A_1} & = & (1-t_{0})(1-t_{1})(1-t_{2})
     (1-t_{01})(1-t_{02})(1-t_{12}) \ceq
     (1-t_{001})(1-t_{002})(1-t_{011})(1-t_{012})
     (1-t_{021})(1-t_{022})(1-t_{112}) \ceq (1-t_{122})
     (1-t_{0001})(1-t_{0002})(1-t_{0011})(1-t_{0012})(1-t_{0021})
            \dots \continue
 &= &  1 - t_0 - t_1 - t_2
 - (t_{01}- t_0 t_1) - (t_{02} - t_0 t_2) - (t_{12} - t_1 t_2) \ceq
- (t_{001} - t_0 t_{01}) - (t_{002} - t_0 t_{02}) - (t_{011} - t_1 t_{01}) \ceq
- (t_{022} - t_2 t_{02}) - (t_{112} - t_1 t_{12}) - (t_{122} - t_2 t_{12}) \ceq
- (t_{012} + t_{021} + t_0 t_1 t_2 - t_0 t_{12} - t_1 t_{02} - t_2 t_{01})
\dots
\label{zc4vA1}
\eea
(for typographical convenience, $\underline{1}$ is replaced by $0$ in
the remainder of this section).
For one-dimensional representations, the characters can be read off 
\cite{DR} the symbol strings:
$
\chi_{A_2}({\bf h_{\tilde{p}} }) = (-1)^{n_0}
$,
$
\chi_{B_1}({\bf h_{\tilde{p}} }) = (-1)^{n_1 }
$,
$
\chi_{B_2}({\bf h_{\tilde{p}} }) = (-1)^{n_0+n_1 }
$,
where $n_{0}$ and $n_1$ are the number of times symbols ${0}$, $1$ appear in
the $\tilde{p}$ symbol string.
For $B_2$ all $t_p$ with an odd total number of $0$'s and $1$'s change sign:
\bea
1/\zeta_{B_2} & = & (1+t_{0})(1+t_{1})(1-t_{2})
     (1-t_{01})(1+t_{02})(1+t_{12}) \ceq
     (1+t_{001})(1-t_{002})(1+t_{011})(1-t_{012})
     (1-t_{021})(1+t_{022})(1-t_{112}) \ceq (1+t_{122})
     (1-t_{0001})(1+t_{0002})(1-t_{0011})(1+t_{0012})(1+t_{0021})
            \dots \continue
 &= &  1 + t_0 + t_1 - t_2
 - (t_{01}- t_0 t_1) + (t_{02} - t_0 t_2) + (t_{12} - t_1 t_2) \ceq
+ (t_{001} - t_0 t_{01}) - (t_{002} - t_0 t_{02}) + (t_{011} - t_1 t_{01}) \ceq
+ (t_{022} - t_2 t_{02}) - (t_{112} - t_1 t_{12}) + (t_{122} - t_2 t_{12}) \ceq
- (t_{012} + t_{021} + t_0 t_1 t_2 - t_0 t_{12} - t_1 t_{02} - t_2 t_{01})
\dots
\label{zc4vB2}
\eea
The form of the remaining cycle expansions depends
crucially on the special role played by the boundary
orbits: by (\ref{bounda4}) the orbit $t_2$ does not contribute
to $A_2$ and $B_1$,
\bea
1/\zeta_{A_2} & = & (1+t_{0})(1-t_{1})
     (1+t_{01})(1+t_{02})(1-t_{12}) \ceq
     (1-t_{001})(1-t_{002})(1+t_{011})(1+t_{012})
     (1+t_{021})(1+t_{022})(1-t_{112}) \ceq (1-t_{122})
     (1+t_{0001})(1+t_{0002})(1-t_{0011})(1-t_{0012})(1-t_{0021})
            \dots \continue
 &= &  1 + t_0 - t_1
 + (t_{01} - t_0 t_1) + t_{02} - t_{12} \ceq
 - (t_{001} - t_0 t_{01}) - (t_{002} - t_0 t_{02}) + (t_{011} + t_1 t_{01}) \ceq
 + t_{022} - t_{122} - (t_{112} - t_1 t_{12})
 + (t_{012} + t_{021} - t_0 t_{12} - t_1 t_{02})
\dots
\label{z4vA2}
\eea
and
\bea
1/\zeta_{B_1} & = & (1-t_{0})(1+t_{1})
     (1+t_{01})(1-t_{02})(1+t_{12}) \ceq
     (1+t_{001})(1-t_{002})(1-t_{011})(1+t_{012})
     (1+t_{021})(1-t_{022})(1-t_{112}) \ceq (1+t_{122})
     (1+t_{0001})(1-t_{0002})(1-t_{0011})(1+t_{0012})(1+t_{0021})
            \dots \continue
 &= &  1 - t_0 + t_1
 + (t_{01} - t_0 t_1) - t_{02} + t_{12} \ceq
 + (t_{001} - t_0 t_{01}) - (t_{002} - t_0 t_{02}) - (t_{011} - t_1 t_{01}) \ceq
 - t_{022} + t_{122} - (t_{112} - t_1 t_{12})
 + (t_{012} + t_{021} - t_0 t_{12} - t_1 t_{02})
\dots
\label{z4vB1}
\eea
In the above we have  assumed that $t_2$ does not change sign under $C_{4v}$
reflections.
For the mixed-symmetry subspace $E$ the curvature expansion is given by
\bea
1/\zeta_E &=&  1 + t_2  + ( -{{t_{0}}^2} + {{t_{1}}^2} )  +
   ( 2 t_{002} - t_2 {{t_{0}}^2} - 2 t_{112} + t_2 {{t_{1}}^2} )
  \ceq
   + ( 2 t_{0011} - 2 t_{0022} + 2 t_2 t_{002} - {{t_{01}}^2} -
      {{t_{02}}^2} + 2 t_{1122} - 2 t_2 t_{112}
  \ceq
 + {{t_{12}}^2} -  {{t_{0}}^2} {{t_{1}}^2} )
   +  ( 2 t_{00002} - 2 t_{00112} + 2 t_2 t_{0011} - 2 t_{00121}
      -  2 t_{00211}
  \ceq
 + 2 t_{00222} - 2 t_2 t_{0022} + 2 t_{01012}
 + 2 t_{01021} - 2 t_{01102}
 - t_2 {{t_{01}}^2} + 2 t_{02022}
            \ceq
   -    t_2 {{t_{02}}^2} + 2 t_{11112} - 2 t_{11222} + 2 t_2 t_{1122} -
      2 t_{12122} + t_2 {{t_{12}}^2}
 - t_2 {{t_{0}}^2} {{t_{1}}^2}
            \ceq
+  2 t_{002} ( -{{t_{0}}^2} + {{t_{1}}^2} )  -
      2 t_{112} ( -{{t_{0}}^2} + {{t_{1}}^2} )  )
\eea

A quick test of the
$\zeta= \zeta_{A_1} \zeta_{A_2} \zeta_{B_1} \zeta_{B_2} \zeta_E^2 $
factorization is afforded by the topological polynomial; substituting
$t_p = z^{n_p}$ into the expansion yields
\[
1/\zeta_{A_1} = 1- 3 z \,\, , \quad
1/\zeta_{A_2}  = 1/\zeta_{B_1}  = 1 \,\, , \quad
1/\zeta_{B_2} = 1/\zeta_{E} = 1 +  z \,\, ,
\]\noindent
in agreement with (\ref{4disk_top}).

\subsection{$C_{2v}$ factorization}
\label{c2vinv}

An arrangement of four identical disks on the vertices of a rectangle has
$C_{2v}$ symmetry (fig.\ref{fgC2v}). $C_{2v}$ consists of
$\{ {e}, {\sigma}_x, {\sigma}_y, {C}_2\}$, {\em i.e.},  the reflections
across the symmetry axes and a rotation by $\pi$. $C_{2v}$ is the symmetry of
several systems studied in the literature, such as the stadium
billiard\rf{bunim}, and the 2-dimensional
anisotropic Kepler potential\rf{gutz_sym}.

This system affords a rather easy visualisation of the  conversion of a 4-disk
dynamics into a fundamental domain symbolic dynamics. An orbit leaving the
fundamental domain through one of the axis may be folded back by a
reflection on that axis; with these symmetry operations
$g_0 = \sigma_x$ and $g_1 = \sigma_y$ we associate labels $1$ and $0$,
respectively. Orbits going to the diagonally opposed disk cross the
boundaries of the fundamental domain twice; the product of these
two reflections is just $C_2=\sigma_x \sigma_y$, to which we assign
the label $2$. For example, a ternary string $0\,0\,1\,0\,2\,0\,1\dots$
is converted into 12143123$\dots$, and the associated group-theory weight is
given by $\dots g_1 g_0 g_2 g_0 g_1 g_0 g_0 $.

Short ternary cycles and the corresponding 4-disk cycles are listed in table~7.
Note that already at length three there is a pair of
cycles (012~=~143 and 021~=~142) related by time
reversal, but {\em not} by any $C_{2v}$ symmetries.

The above is the complete description of the symbolic dynamics for 4
sufficiently separated equal
disks placed at corners of a rectangle. However, if
the fundamental domain requires further partitioning, the ternary description is
insufficient. For example, in the stadium billiard fundamental domain one has
to distinguish between bounces off the straight and the curved sections of the
billiard wall; in that case there exists evidence\rf{Biham} that five
symbols suffice for constructing the covering symbolic dynamics.

The group $C_{2v}$ has four one-dimensional
representations, distinguished by their behaviour under axis reflections.
The $A_1$ representation is symmetric with respect to both reflections;
the $A_2$ representation is antisymmetric with respect to both.
The $B_1$ and $B_2$ representations are symmetric under one
and antisymmetric under the other reflection. The character table is
\vskip .3cm
\begin{center}
\begin{tabular}{|c|crrr|}
\hline
$C_{2v}$    & $A_1$& $A_2$& $B_1$ & $B_2$\\
\hline
$ e   $       & $1$  & $ 1$  & $ 1$  & $ 1$  \\
$ C_2$        & $1$  & $ 1$  & $-1$  & $-1$  \\
$ \sigma_x$   & $1$  & $-1$  & $ 1$  & $-1$  \\
$ \sigma_{y} $& $1$  & $-1$  & $-1$  & $ 1$  \\
\hline
\end{tabular}
\end{center}
\vskip .3cm

Substituted into the factorized determinant (\ref{factzet}),
the contributions of periodic orbits split as follows
\vskip 12pt
\begin{tabular}{rlccccc}

$g_{\tilde p}$
&  & &  $A_1$  &  $A_2$  &  $B_1$  &  $B_2$  \\
$e$:
& $(1-t_{\tilde p} )^4 $ &=&  $(1-t_{\tilde p})$ & $(1-t_{\tilde p})$ &
                        $(1-t_{\tilde p})$ &$(1-t_{\tilde p})$ \\
$C_2$:
& $(1-t_{\tilde p}^2 )^2$ &=&  $(1-t_{\tilde p})$ & $(1-t_{\tilde p})$ &
                        $(1-t_{\tilde p})$ &$(1-t_{\tilde p})$  \\
$\sigma_x$:
& $(1-t_{\tilde p}^2 )^2$&=& $(1-t_{\tilde p})$ & $(1+t_{\tilde p})$ &
                        $(1-t_{\tilde p})$ &$(1+t_{\tilde p})$ \\
$\sigma_{y}$:
& $(1-t_{\tilde p}^2 )^2$&=& $(1-t_{\tilde p})$ & $(1+t_{\tilde p})$ &
                        $(1+t_{\tilde p})$ &$(1-t_{\tilde p})$ \\
\end{tabular}
\vskip 12pt
\noindent
Cycle expansions follow by substituting cycles and their group
theory factors from table~7.
For $A_1$ all characters are $+1$, and the corresponding cycle expansion is
given in (\ref{zc4vA1}).
Similarly, the totally antisymmetric subspace factorization $A_2$ is given by
(\ref{zc4vB2}),
the $B_2$ factorization of $C_{4v}$.
For $B_1$ all $t_p$ with an odd total number of $0$'s and $2$'s change sign:
\bea
1/\zeta_{B_1} & = & (1+t_{0})(1-t_{1})(1+t_{2})
     (1+t_{01})(1-t_{02})(1+t_{12}) \ceq
     (1-t_{001})(1+t_{002})(1+t_{011})(1-t_{012})
     (1-t_{021})(1+t_{022})(1+t_{112}) \ceq
     (1-t_{122})
     (1+t_{0001})(1-t_{0002})(1-t_{0011})(1+t_{0012})(1+t_{0021})
            \dots \continue
 &= &  1 + t_0 - t_1 + t_2
 + (t_{01}- t_0 t_1) - (t_{02} - t_0 t_2) + (t_{12} - t_1 t_2) \ceq
- (t_{001} - t_0 t_{01}) + (t_{002} - t_0 t_{02}) + (t_{011} - t_1 t_{01}) \ceq
+ (t_{022} - t_2 t_{02}) + (t_{112} - t_1 t_{12}) - (t_{122} - t_2 t_{12}) \ceq
- (t_{012} + t_{021} + t_0 t_1 t_2 - t_0 t_{12} - t_1 t_{02} - t_2 t_{01})
\dots
\label{zc2vB1}
\eea
For $B_2$ all $t_p$ with an odd total number of $1$'s and $2$'s change sign:
\bea
1/\zeta_{B_2} & = & (1-t_{0})(1+t_{1})(1+t_{2})
     (1+t_{01})(1+t_{02})(1-t_{12}) \ceq
     (1+t_{001})(1+t_{002})(1-t_{011})(1-t_{012})
     (1-t_{021})(1-t_{022})(1+t_{112}) \ceq
     (1+t_{122})
     (1+t_{0001})(1+t_{0002})(1-t_{0011})(1-t_{0012})(1-t_{0021})
            \dots \continue
 &= &  1 - t_0 + t_1 + t_2
+ (t_{01}- t_0 t_1) + (t_{02} - t_0 t_2) - (t_{12} - t_1 t_2) \ceq
+ (t_{001} - t_0 t_{01}) + (t_{002} - t_0 t_{02}) - (t_{011} - t_1 t_{01}) \ceq
- (t_{022} - t_2 t_{02}) + (t_{112} - t_1 t_{12}) + (t_{122} - t_2 t_{12}) \ceq
- (t_{012} + t_{021} + t_0 t_1 t_2 - t_0 t_{12} - t_1 t_{02} - t_2 t_{01})
\dots
\label{zc2vB2}
\eea
Note that all of the above cycle expansions group long orbits together
with their pseudo-orbit shadows,
so that the shadowing arguments for convergence\rf{AACI} still apply.

The topological polynomial factorizes as
\[
{1 \over \zeta_{A_1}} = 1-3z \quad,\quad
{1 \over \zeta_{A_2}} =
{1 \over \zeta_{B_1}} =
{1 \over \zeta_{B_2}} = 1 + z,
\]\noindent
consistent with the 4-disk factorization (\ref{4disk_top}).

\section{Summary}
\label{Summar}

The techniques of this paper have been applied to computations of the
3-disk classical and quantum spectra in refs.~\cite{CE,CES,PS}, and to a
``Zeeman effect" pinball and the $x^2 y^2$ potentials in refs.~\cite{russ,DR}.
The main lesson of such calculations is that if a dynamical system has a
discrete symmetry, the symmetry should be exploited; much is gained, both in
understanding of the spectra and ease of their evaluation. Once this is
appreciated, it is hard to conceive of a calculation without factorization; it
would correspond to quantum mechanical calculations without wave--function
symmetrizations.

In a larger perspective, the factorizations developed above are special cases
of a general approach to exploiting the group-theoretic invariances
in spectra computations, such as those used in enumeration of periodic
geodesics\rf{BV,mcke,steiner} for hyperbolic billiards\rf{Had,gutz_hyp}
and Selberg zeta functions\rf{gutbook}.

Reduction to the fundamental domain simplifies symbolic dynamics and
eliminates symmetry induced degeneracies. While the
resummation of the theory from the trace sums\rf{gutz} to
the cycle expansions\rf{cycprl}
does not reduce the exponential growth in number of cycles with the cycle
length, in practice only the short orbits are used, and for them the labor
saving is dramatic. For example, for the 3-disk pinball there are $256$
periodic points of length $8$, but reduction to the fundamental domain
non-degenerate prime cycles reduces the number of the distinct cycles of length
8 to 30.

In addition, as we demonstrate by explicit calculations in ref.~\cite{eck,CES},
cycle expansions of the symmetry reduced zeta functions converge dramatically
faster than the unfactorized zeta functions. One reason is that the
unfactorized zeta function has many closely spaced zeros and zeros of
multiplicity higher than one; since the cycle expansion is a polynomial
expansion in topological cycle length, accomodating such behaviour requires
many terms. The zeta functions on separate subspaces have more evenly
and widely spaced zeros, are smoother, do not have symmetry-induced multiple
zeros, and fewer cycle expansion terms (short cycle truncations) suffice to
determine them.
Furthermore, the cycles in the fundamental domain sample
phase space more densely than in the full space. For example, for the 3-disk
problem, there are 9 distinct (symmetry unrelated) cycles
of length $7$ or less in full space, corresponding to $47$ distinct
periodic points. In the fundamental domain, we have $8$ (distinct) periodic
orbits up to length $4$ and thus $22$ different periodic points in $1/6$-th
the phase space,  {\em i.e.}, an increase in density by a factor $3$ with
the same numerical effort. 

We emphasize that the symmetry factorization (\ref{symm}) of the dynamical
zeta function is {\em intrinsic} to the classical dynamics, and not a special
property of quantal spectra (in which context it was used before\rf{gasp}).
Our results are not restricted to the Hamiltonian systems, or only to the
configuration space symmetries; for example, the discrete symmetry can be a
symmetry of the Hamiltonian phase space\rf{robb}.
In conclusion, the manifold advantages of the
symmetry reduced dynamics should thus be obvious; full
space cycle expansions, such as those included in the appendix,
are useful only for cross checking purposes.

\section{Appendix}
\label{Append}

Here  we list the 3- and 4-disk cycle expansions for unfactorized zeta
functions. They are not recommended for actual computations, as the factorized
zeta functions yield much better spectra, but they might be useful for
cross-checking purposes.

For the 3-disk pinball (assuming no symmetries between
disks) the curvature expansion (\ref{det(1-L)}) is given by:
\bea
1/\zeta & = & (1-z^2t_{12})(1-z^2t_{13})(1-z^2t_{23})
     (1-z^3t_{123})(1-z^3t_{132})
        \ceq
     (1-z^4t_{1213})(1-z^4t_{1232})(1-z^4t_{1323})(1-z^5t_{12123})\cdots
        \continue
     & = & 1 - z^2t_{12} - z^2t_{23} - z^2 t_{31}
     -  z^3 t_{123} -  z^3t_{132}
 \ceq
     -  z^4 \lbrack (t_{1213} - t_{12}t_{13})
     +  (t_{1232} - t_{12}t_{23})
     +  (t_{1323} - t_{13}t_{23}) \rbrack
         \ceq
     -  z^5 \lbrack ( t_{12123} - t_{12} t_{123}) + \cdots\rbrack - \cdots
\label{3dz}
\eea

The symmetrically arranged 3-disk pinball cycle expansion of
the Euler product (\ref{eul_sym}) (see table~2 and fig.\ref{fgC3v}) is given by:
\bea
1/\zeta & = & (1-z^2t_{12})^3 (1-z^3t_{123})^2 (1-z^4t_{1213})^3 \ceq
    (1-z^5t_{12123})^6 (1-z^6t_{121213})^{6}(1-z^6t_{121323})^3  \dots
        \continue
     & = & 1 - 3z^2\, t_{12} - 2z^3\, t_{123} - 3z^4\, (t_{1213} - t_{12}^2)
     - 6z^5\, ( t_{12123} - t_{12} t_{123}) \ceq
     - z^6\,( 6\, t_{121213} + 3\, t_{121323} +  t_{12}^3
                - 9\, t_{12} t_{1213} - t_{123}^2 )  \ceq
     - 6z^7\,( t_{1212123} + t_{1212313} + t_{1213123}
               + t_{12}^2 t_{123}  - 3\, t_{12} t_{12123}
               - t_{123} t_{1213}  ) \ceq
     -  3z^8\,( 2\, t_{12121213} + t_{12121313}
                  + 2\, t_{12121323} + 2\, t_{12123123} \ceq
                 ~~~~~ + 2\, t_{12123213} + t_{12132123}
        + 3\, t_{12}^2 t_{1213} + t_{12} t_{123}^2 \ceq
       ~~~~~- 6\, t_{12} t_{121213} - 3\, t_{12} t_{121323} -
4\, t_{123} t_{12123} - t_{1213}^2) - \cdots
\label{3dzeta}
\eea

For the symmetriclly arranged
 4-disk pinball the symmetry group is C$_{4v}$, of order 8.
The degenerate cycles can have multiplicities 2, 4  or 8
(see table~3):
\bea
1/\zeta & = & (1-z^2t_{12})^4 (1-z^2t_{13})^2 (1-z^3t_{123})^8
     (1-z^4t_{1213})^8
     (1-z^4t_{1214})^4 \ceq
     (1-z^4t_{1234})^2
     (1-z^4t_{1243})^4
     (1-z^5t_{12123})^8
     (1-z^5t_{12124})^8
     (1-z^5t_{12134})^8
       \ceq
     (1-z^5t_{12143})^8
     (1-z^5t_{12313})^8
     (1-z^5t_{12413})^8
       \cdots
\label{4dzeta}
\eea
and the cycle expansion is given by
\bea
1/\zeta & = & 1 - z^2 (4 \,t_{12} + 2 \,t_{13})- 8 z^3 \,t_{123}
       \ceq
     - z^4 ( 8 \,t_{1213} + 4 \,t_{1214} + 2 \,t_{1234} + 4 \,t_{1243}
      -6 \,t_{12}^2 - t_{13}^2 - 8 \,t_{12} t_{13})
       \ceq
     - 8 z^5 ( t_{12123} + t_{12124} + t_{12134}
       + t_{12143} + t_{12313}  + t_{12413}
- 4 \,t_{12} t_{123} - 2 \,t_{13} t_{123})
       \ceq
     - 4 z^6 (2 \,S_8 + \, S_4 +
  \,t_{12}^3 + 3 \,t_{12}^2 \,t_{13} + \,t_{12} t_{13}^2
            - 8 \,t_{12} t_{1213} - 4 \,t_{12} t_{1214}  \ceq
       - 2 \,t_{12} t_{1234}  - 4 \,t_{12} t_{1243}
      - 4 \,t_{13} t_{1213} - 2 \,t_{13} t_{1214} - \,t_{13} t_{1234}
   \ceq
       - 2 \,t_{13} t_{1243} - 7 \,t_{123}^2 ) - \cdots
\eea
where in the coefficient to $z^6$ the abbreviations $S_8$ and $S_4$
stand for the sums over the weights of the $12$ orbits with multiplicity $8$
and the $5$ orbits of multiplicity $4$, respectively; the orbits are listed
in table 2.

\vskip 20pt
{\bf Acknowledgements}

We are grateful to the hospitality of the Max Planck Institut f\"ur
Mathematik, during whose spring 1988 Chaos and Fractals workshop this
collaboration was initiated. We acknowledge stimulating exchanges with
F.~Christiansen,
P.~Dahlqvist,
G.~Eilenberger,
P.~Grassberger,
M.~Gutzwiller,
K.T.~Hansen,
T.~Janssen,
B.~Lauritzen,
G.~Ott,
J.M.~Robbins,
H.H.~Rugh,
G.~Russberg,
M.~Sieber,
and D.~Wintgen.
P.C. is grateful to the Carlsberg Fundation for support, to I.
Procaccia of the Weizmann Institute, P. Hemmer of University of Trondheim and
J. Lowenstein of New York University, where parts of this work were done,
for the hospitality. B.E. was supported in part by the National Science
Foundation under Grant No. PHY82-17853, supplemented by funds from the National
Aeronautics and Space Administration.

\newpage

\newcommand{\AP}[1]{{\em Ann.\ Phys.}\/ {\bf #1}}
\newcommand{\CM}[1]{{\em Cont.\ Math.}\/ {\bf #1}}
\newcommand{\CMP}[1]{{\em Commun.\ Math.\ Phys.}\/ {\bf #1}}
\newcommand{\INCB}[1]{{\em Il Nuov.\ Cim.\ B}\/ {\bf #1}}
\newcommand{\JCP}[1]{{\em J.\ Chem.\ Phys.}\/ {\bf #1}}
\newcommand{\JETP}[1]{{\em Sov.\ Phys.\ JETP}\/ {\bf #1}}
\newcommand{\JETPL}[1]{{\em JETP Lett.\ }\/ {\bf #1}}
\newcommand{\JMP}[1]{{\em J.\ Math.\ Phys.}\/ {\bf #1}}
\newcommand{\JMPA}[1]{{\em J.\ Math.\ Pure Appl.}\/ {\bf #1}}
\newcommand{\JPA}[1]{{\em J.\ Phys.}\/ {\bf A  #1}}
\newcommand{\JPB}[1]{{\em J.\ Phys.}\/ {\bf B  #1}}
\newcommand{\JPC}[1]{{\em J.\ Phys.\ Chem.}\/ {\bf #1}}
\newcommand{\NPB}[1]{{\em Nucl.\ Phys.}\/ {\bf B #1}}
\newcommand{\PLA}[1]{{\em Phys.\ Lett.}\/ {\bf A #1}}
\newcommand{\PLB}[1]{{\em Phys.\ Lett.}\/ {\bf B #1}}
\newcommand{\PRA}[1]{{\em Phys.\ Rev.}\/ {\bf A #1}}
\newcommand{\PRD}[1]{{\em Phys.\ Rev.}\/ {\bf D #1}}
\newcommand{\PRL}[1]{{\em Phys.\ Rev.\ Lett.}\/ {\bf #1}}
\newcommand{\PST}[1]{{\em Phys.\ Scripta }\/ {\bf T #1}}
\newcommand{\RMS}[1]{{\em Russ.\ Math.\ Surv.}\/ {\bf #1}}
\newcommand{\USSR}[1]{{\em Math.\ USSR.\ Sb.}\/ {\bf #1}}
\renewcommand{\baselinestretch} {1}

\newpage

{\bf {Figure captions}}
\vskip 20pt

\begin{enumerate}

\fg{fgC3v}
The symmetries of three disks on an equilateral triangle.
The fundamental domain is indicated by the shaded wedge.

\fg{fg3_exmp}
Some examples of 3-disk cycles.
(a) $\overline{12123}$ and  $\overline{13132}$ are
mapped into each other by $\sigma_{23}$, the flip across 1 axis;
this orbit has degeneracy 6 under $C_{3v}$ symmetries.
Similarly (b) $\overline{123}$ and $\overline{132}$ and
(c) $\overline{1213}$, $\overline{1232}$ and $\overline{1323}$
are degenerate under $C_{3v}$.
The orbits (d) $\overline{121212313}$ and $\overline{121212323}$
are related by time reversal but not by any $C_{3v}$ symmetry.

\fg{fg3_dsk}
The scattering geometry for the disk
radius/separation ratio $a:R = 1:2.5$. (a) the three disks,
with $\overline{12}$, $\overline{123}$ and
$\overline{121232313}$ cycles indicated.
(b) the fundamental domain, $ie.$ a wedge consisting
of a section of a disk, two segments of symmetry axes acting as
straight mirror walls, and an escape gap. The above cycles
restricted to the fundamental domain are now
the two fixed points $\overline{0}$ and $\overline{1}$ and
the $\overline{100}$ cycle.

\fg{fgC4v}
The symmetries of four disks on a square.
The fundamental domain is indicated by the shaded wedge.

\fg{fgC2v}
The symmetries of four disks on a rectangle.
The fundamental domain is indicated by the shaded wedge.

\end{enumerate}

\vfill\eject
{\small{

{\small
\renewcommand{\arraystretch}{0.8}
\vskip .3cm
\begin{tabular}{|r|c|l|r|l|}
 \hline
n  & $M_n$ & $N_n$   &   $S_n$  &    $m_p \cdot \hat{p}$ \\
\hline
1  &   0  &   0             & 0 &       \\
2  &   3  &   6=3$\cdot$2         & 1 & 3$\cdot$12 \\
3  &   2  &   6=2$\cdot$3         & 1 & 2$\cdot$123 \\
4  &   3  &  18=3$\cdot$2+3$\cdot$4     & 1 & 3$\cdot$1213 \\
5  &   6  &  30=6$\cdot$5         & 1 & 6$\cdot$12123 \\
6  &   9  &  66=3$\cdot$2+2$\cdot$3+9$\cdot$6 & 2
                           & 6$\cdot$121213 + 3$\cdot$121323 \\
7  &  18  & 126=18$\cdot$7        & 3
                & 6$\cdot$1212123 + 6$\cdot$1212313 + 6$\cdot$1213123 \\
8  &  30  & 258=3$\cdot$2+3$\cdot$4+30$\cdot$8& 6 &
        6$\cdot$12121213 + 3$\cdot$12121313 + 6$\cdot$12121323 \\
   &    &    &   &  + 6$\cdot$12123123 + 6$\cdot$12123213 + 3$\cdot$12132123 \\
9  &  56  & 510=2$\cdot$3+56$\cdot$9 & 10 &
        6$\cdot$121212123 + 6$\cdot$(121212313 + 121212323) \\
 & & & & + 6$\cdot$(121213123 + 121213213) + 6$\cdot$121231323 \\
 & & & & + 6$\cdot$(121231213 + 121232123) + 2$\cdot$121232313 \\
 & & & & + 6$\cdot$121321323\\
10  &   99  &   1022             & 18 &       \\
\hline
\end{tabular}
\vskip .3cm

Table 1. List of the 3-disk prime cycles up to length 10. Here $n$ is the
cycle length, $M_n$ the number of prime cycles, $N_n$ the number of periodic
points and $S_n$ the number of distinct prime cycles under the $C_{3v}$
symmetry.
Column~3 also indicates the splitting of $N_n$ into contributions from orbits
of lengths that divide $n$.
The prefactors in the fifth column indicate the degeneracy $m_p$ of
the cycle; for example, 3$\cdot$12 stands for the three prime cycles
$\overline{12}$, $\overline{13}$ and $\overline{23}$ related by $2\pi/3$
rotations. Among symmetry related cycles, a representative $\hat{p}$
which is lexically lowest was chosen. The cycles of length 9 grouped by
parenthesis are related by time reversal symmetry (but not by any other
$C_{3v}$ transformation).

\vskip .5cm

\begin{tabular}{|r|r|l|r|l|}
 \hline
n  & $M_n$& $N_n$   &   $S_n$  &   $m_p \cdot \hat{p}$  \\
\hline
1  &   0  &   0     &        0 &       \\
2  &   6  &  12=6$\cdot$2  & 2 & 4$\cdot$12  +  2$\cdot$13 \\
3  &   8  &  24=8$\cdot$3  & 1 & 8$\cdot$123 \\
4  &  18  &  84=6$\cdot$2+18$\cdot$4  &  4 & 8$\cdot$1213 + 4$\cdot$1214
                                      +  2$\cdot$1234  +  4$\cdot$1243  \\
5&48&240=48$\cdot$5  & 6 & 8$\cdot$(12123 + 12124) + 8$\cdot$12313 \\
 &  &                &   & + 8$\cdot$(12134 + 12143)  + 8$\cdot$12413 \\
6  & 116  & 732=6$\cdot$2+8$\cdot$3+116$\cdot$6 & 17 &
         8$\cdot$121213  +  8$\cdot$121214  +  8$\cdot$121234\\
& & & &  + 8$\cdot$121243  +  8$\cdot$121313  +  8$\cdot$121314\\
& & & &  + 4$\cdot$121323  +  8$\cdot$(121324  +  121423)\\
& & & &  + 4$\cdot$121343  +  8$\cdot$121424  +  4$\cdot$121434\\
& & & &  + 8$\cdot$123124  +  8$\cdot$123134  +  4$\cdot$123143\\
& & & &  + 4$\cdot$124213  +  8$\cdot$124243\\
7  &   312  &   2184             &  39  &       \\
8  &   810  &   6564             & 108  &       \\
\hline
\end{tabular}
\vskip .5cm

Table 2. List of the 4-disk prime cycles up to length 8.
The meaning of the symbols is the same as in Table 1.
Orbits related by time reversal symmetry (but no other symmetry)
already appear at cycle length 5.
List of the cycles of length 7 and 8 has been omitted.

\newpage

\begin{tabular}{|r|c|rrr|rr|}
\hline
n  & $M_n(N)$& $M_n(2)$& $M_n(3)$& $M_n(4)$
\cr \hline
    1  &   N                     &  2 &    3 &      4 
\cr 2  &$N (N-1)/2$              &  1 &    3 &      6 
\cr 3  &$N (N^2-1)/3$            &  2 &    8 &     20 
\cr 4  &$N^2(N^2-1)/4$           &  3 &   18 &     60 
\cr 5  &$(N^5-N)/5$              &  6 &   48 &    204 
\cr 6  &$(N^6-N^3-N^2+N)/ 6$     &  9 &  116 &    670 
\cr 7  &$(N^7-N)/ 7$             & 18 &  312 &   2340 
\cr 8  &$N^4(N^4-1)/ 8$          & 30 &  810 &   8160 
\cr 9  &$N^3(N^6-1)/ 9$          & 56 & 2184 &  29120 
\cr 10 &$(N^{10}-N^5-N^2+ N)/ 10$& 99 & 5880 & 104754 
\cr \hline
\end{tabular}
\vskip .3cm

Table 3. Number of prime cycles for various alphabets and grammars up to
length 10. The first column gives the cycle length, the second the formula
(\ref{Moeb_inv}) for the number of prime cycles for complete $N$-symbol
dynamics, columns three through five give the numbers for $N=2, 3$ and $4$.

\vskip .8cm

\begin{tabular}{|l|l|l|}
\hline
$\tilde{p}$ & ${p}$  & $m_p$ \\
\hline
 1  & $+$ & 2 \\
 0 & $-+$ & 1 \\ \hline
 01 & $--\;++$   & 1 \\ \hline
 001 & $-++$  & 2 \\
 011 & $---\;+++$ & 1 \\
\hline
 0001 & $-+--\;+-++$  & 1 \\
 0011 & $-+++$ & 2 \\
 0111 & $----\;++++$ & 1 \\
\hline
 00001 & $-+-+-$ & 2 \\
 00011 & $-+---\;+-+++$  & $1$ \\
 00101 & $-++--\;+--++$ & $1$ \\
 00111 & $-+---\;+-+++$   &    $1$ \\
 01011 & $--+++$   &    $2$ \\
 01111 & $-----\;+++++$ & $1$ \\
\hline
 001011 & $-++---\;+--+++$ & $1$ \\
 001101 & $-+++--\;+---++$ & $1$ \\
\hline
\end{tabular}

\vskip .5cm

Table 4.
Correspondence between the $C_2$ symmetry reduced cycles $\tilde{p}$ and the
standard Ising model periodic configurations ${p}$, together with their
multiplicities $m_p$. Also listed are the two shortest cycles (length 6)
related by time reversal, but distinct under $C_2$.

\newpage

\begin{tabular}{|l|l|l|}
\hline
$\tilde{p}$ & ${p}$  & ${\bf g}_{\tilde{p}}$ \\
\hline
0  &  1\,2  &  ${\sigma}_{12}$ \\
1  &  1\,2\,3  &  $C_3$ \\
\hline
01  &  12\,13  &  ${\sigma}_{23}$ \\
\hline
001  &  121\,232\,313  &  $C_3$ \\
011  &  121\,323  &  ${\sigma}_{13}$ \\
\hline
0001  &  1212\,1313  &  ${\sigma}_{23}$ \\
0011  &  1212\,3131\,2323  &  $C_3^2$ \\
0111  &  1213\,2123  &  ${\sigma}_{12}$ \\
\hline
00001  &  12121\,23232\,31313  &  $C_3$ \\
00011  &  12121\,32323  &  ${\sigma}_{13}$ \\
00101  &  12123\,21213  &  ${\sigma}_{12}$ \\
00111  &  12123  &  $e$ \\
01011  &  12131\,23212\,31323  &  $C_3$ \\
01111  &  12132\,13123  &  ${\sigma}_{23}$ \\
\hline
000001  &  121212\,131313  &  ${\sigma}_{23}$ \\
000011  &  121212\,313131\,232323  &  $C_3^2$ \\
000101  &  121213  &  $e$ \\
000111  &  121213\,212123  &  ${\sigma}_{12}$ \\
001011  &  121232\,131323  &  ${\sigma}_{23}$ \\
001101  &  121231\,323213  &  ${\sigma}_{13}$ \\
001111  &  121231\,232312\,313123  &  $C_3$ \\
010111  &  121312\,313231\,232123  &  $C_3^2$ \\
011111  &  121321\,323123  &  ${\sigma}_{13}$ \\
\hline
\end{tabular}

\vskip .5cm

Table 5.
$C_{3v}$ correspondence between the binary labelled fundamental domain prime
cycles $\tilde{p}$ and the full 3-disk ternary \{1,2,3\} 
labelled cycles ${p}$, together
with the $C_{3v}$ transformation that maps the end point of the
$\tilde{p}$ cycle into the irreducible segment of the $p$ cycle. The degeneracy
of $p$ cycle is $m_p=6 n_{\tilde{p}}/n_p$. The shortest
pair of the fundamental domain cycles related by time symmetry are the
6-cycles $\overline{001011}$ and $\overline{001101}$.

\newpage

\begin{tabular}{|l|l|l|}
\hline
$\tilde{p}$ & ${p}$  & ${\bf h}_{\tilde{p}}$ \\
\hline
0  &  1\,2  &  ${\sigma}_x$ \\
1  &  1\,2\,3\,4  &  $C_4$ \\
2  &  1\,3  &  $C_2$, ${\sigma}_{13}$ \\
\hline
01  &  12\,14  &  ${\sigma}_{24}$ \\
02  &  12\,43  &  ${\sigma}_y$ \\
12  &  12\,41\,34\,23  &  $C_4^3$ \\
\hline
001  &  121\,232\,343\,414  &  $C_4$ \\
002  &  121\,343  &  $C_2$ \\
011  &  121\,434  &  ${\sigma}_y$ \\
012  &  121\,323  &  ${\sigma}_{13}$ \\
021  &  124\,324  &  ${\sigma}_{13}$ \\
022  &  124\,213  &  ${\sigma}_x$ \\
112  &  123  &  $e$ \\
122  &  124\,231\,342\,413  &  $C_4$ \\
\hline
0001  &  1212\,1414  &  ${\sigma}_{24}$ \\
0002  &  1212\,4343  &  ${\sigma}_y$ \\
0011  &  1212\,3434  &  $C_2$ \\
0012  &  1212\,4141\,3434\,2323 & $C_4^3$~~~$\*$ \\
0021\ $(a)$ &  1213\,4142\,3431\,2324  &  $C_4^3$ \\
0022  &  1213  &  $e$ \\
0102\ $(a)$  &  1214\,2321\,3432\,4143 & $C_4$~~~$\*$ \\
0111  &  1214\,3234  &  ${\sigma}_{13}$ \\
0112\ $(b)$  &  1214\,2123  &  ${\sigma}_x$ \\
0121\ $(b)$  &  1213\,2124  &  ${\sigma}_x$ \\
0122  &  1213\,1413  &  ${\sigma}_{24}$ \\
0211  &  1243\,2134  &  ${\sigma}_x$ \\
0212  &  1243\,1423  &  ${\sigma}_{24}$ \\
0221  &  1242\,1424  &  ${\sigma}_{24}$ \\
0222  &  1242\,4313  &  ${\sigma}_y$ \\
1112  &  1234\,2341\,3412\,4123  &  $C_4$ \\
1122  &  1231\,3413  &  $C_2$ \\
1222  &  1242\,4131\,3424\,2313  &  $C_4^3$ \\
\hline
\end{tabular}
\vskip .5cm

Table 6.
$C_{4v}$ correspondence between the ternary
fundamental
domain prime cycles $\tilde{p}$ and the full 4-disk \{1,2,3,4\} 
labelled cycles ${p}$,
together with the $C_{4v}$ transformation that maps the end point of the
$\tilde{p}$ cycle into an irreducible segment of the $p$ cycle.
For typographical convenience, the symbol $\underline{1}$ 
of sect.~\ref{C_4v_inva} has been replaced
by $0$, so that the ternary alphabet is $\{0,1,2\}$. The degeneracy
of the $p$ cycle is $m_p=8 n_{\tilde{p}}/n_p$.
Orbit $\overline{2}$ is the sole boundary orbit,
invariant both under a rotation by $\pi$
and a reflection across a diagonal. The two pairs of cycles
marked by $(a)$ and $(b)$ are related by time reversal, but cannot
be mapped into each other by $C_{4v}$ transformations.

\newpage

\begin{tabular}{|l|l|l|}
\hline
$\tilde{p}$ & $p$       &       ${ \bf g}$ \\ \hline
0       &       1\,4      &       ${ \sigma}_y$ \\
1       &       1\,2      &       ${ \sigma}_x$ \\
2       &       1\,3      &       ${ C}_2$ \\
\hline
01      &       14\,32    &       ${ C}_2$ \\
02      &       14\,23    &       ${ \sigma}_x$ \\
12      &       12\,43    &       ${ \sigma}_y$ \\
\hline
001     &       141\,232  &       ${ \sigma}_x$ \\
002     &       141\,323  &       ${ C}_2$ \\
011     &       143\,412  &       ${ \sigma}_y$ \\
012     &       143     &       ${ e}$ \\
021     &       142     &       ${ e}$ \\
022     &       142\,413  &       ${ \sigma}_y$ \\
112     &       121\,343  &       ${ C}_2$ \\
122     &       124\,213  &       ${ \sigma}_x$ \\
\hline
0001    &  1414\,3232 & ${C}_2$ \\
0002    &  1414\,2323 & ${\sigma}_x$ \\
0011    &       1412              & ${ e}$ \\
0012    &       1412\,4143        & ${ \sigma}_y$ \\
0021    &       1413\,4142        & ${ \sigma}_y$ \\
0022    &       1413              & ${ e}$ \\
0102    &       1432\,4123        & ${ \sigma}_y$ \\
0111    &       1434\,3212        &       ${ C}_2$ \\
0112    &       1434\,2343        &       ${ \sigma}_x$ \\
0121    &       1431\,2342        &       ${ \sigma}_x$ \\
0122    &       1431\,3213        &       ${ C}_2$ \\
0211    &       1421\,2312        &       ${ \sigma}_x$ \\
0212    &       1421\,3243        &       ${ C}_2$ \\
0221    &       1424\,3242        &       ${ C}_2$ \\
0222    &       1424\,2313        &       ${ \sigma}_x$ \\
1112    &       1212\,4343        &       ${ \sigma}_y$ \\
1122    &       1213    &       ${ e}$ \\
1222    &       1242\,4313        &       ${ \sigma}_y$ \\
\hline
\end{tabular}
\vskip .5cm

Table 7.
$C_{2v}$ correspondence between the ternary $\{0,1,2\}$
fundamental domain prime cycles $\tilde{p}$ and the full 4-disk \{1,2,3,4\}
cycles ${p}$, together with the $C_{2v}$ transformation that maps the end
point of the $\tilde{p}$ cycle into an irreducible segment of the $p$ cycle.
The degeneracy of the $p$ cycle is $m_p=4 n_{\tilde{p}}/n_p$. Note that the
$ 012$ and $ 021 $ cycles are related by time reversal,
but cannot be mapped into each other by $C_{2v}$ transformations. The full
space orbit listed here is generated from the symmetry reduced code by the
rules given in section~\ref{c2vinv}, starting from disk 1.
}

}} 

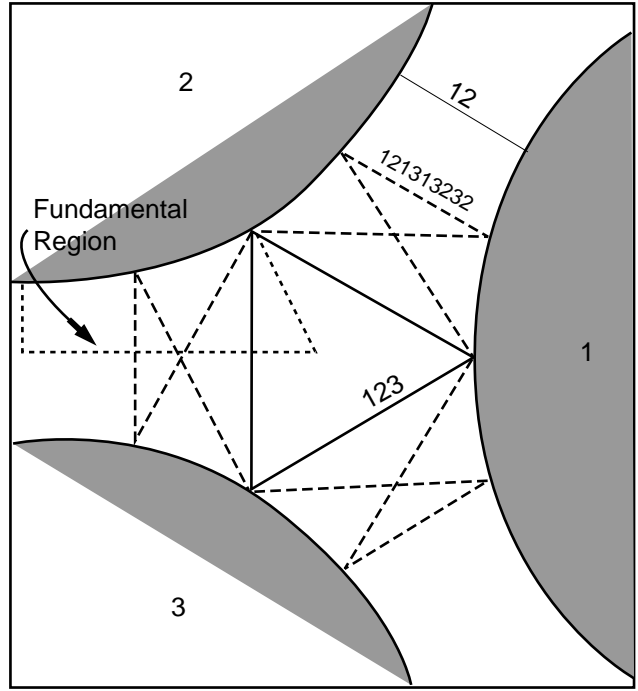
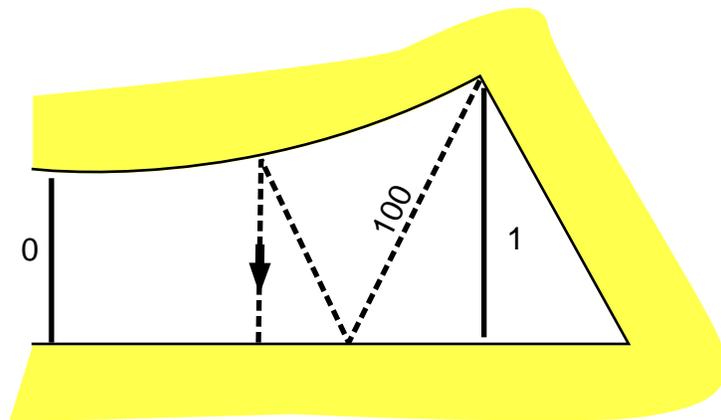
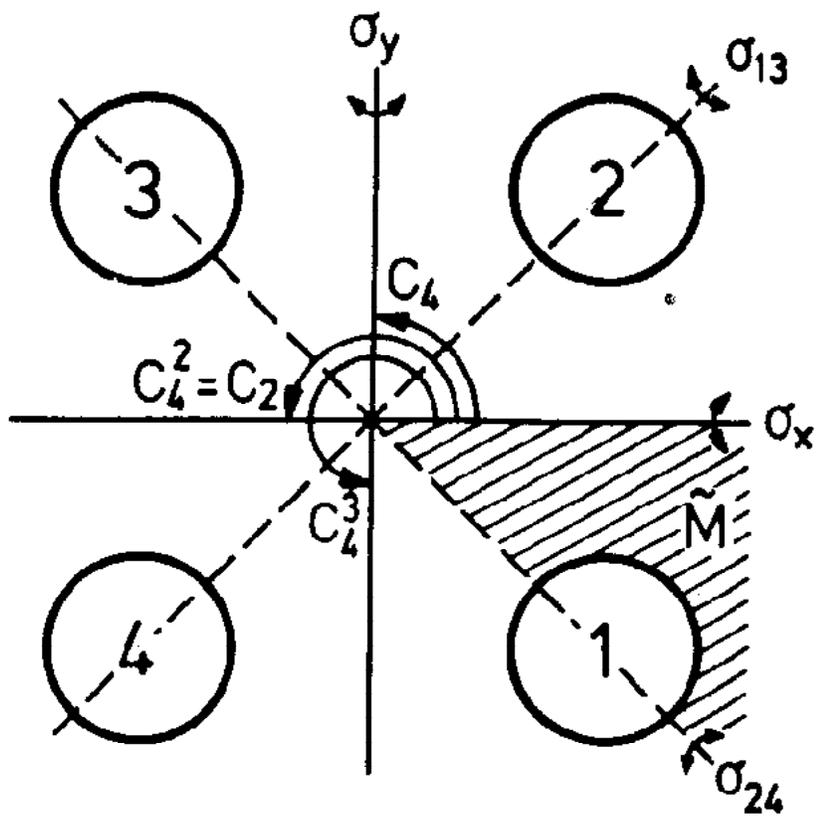
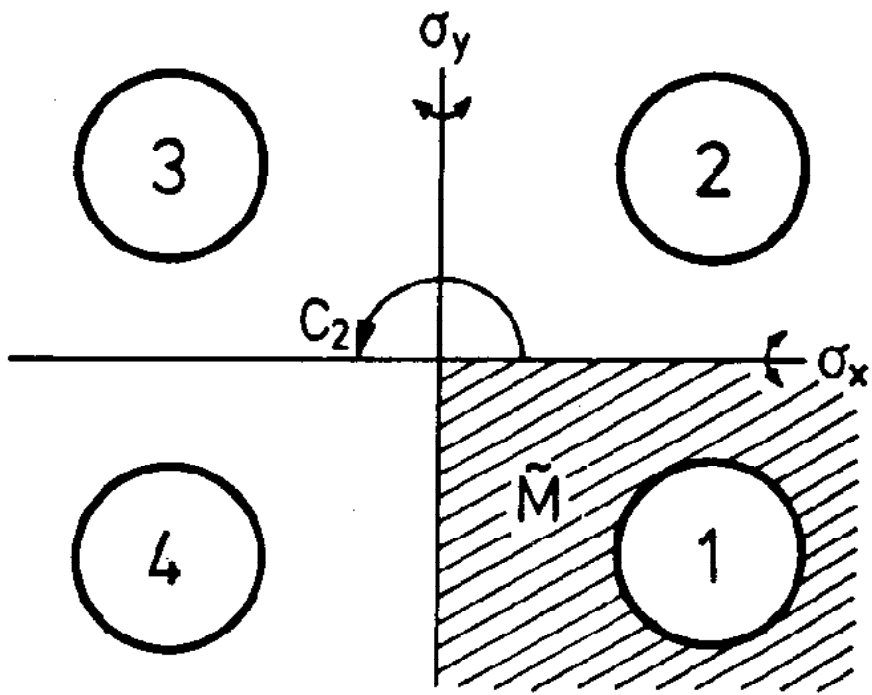
\end{document}